\documentclass[twocolumn,superscriptaddress,
amsmath,amssymb,aps,prl]{revtex4-1}

\usepackage{graphicx}
\usepackage{physics}
\usepackage{comment}
\usepackage{color}
\usepackage{dcolumn}
\usepackage[normalem]{ulem}
\usepackage{bm}
\usepackage{multirow}
\usepackage[T1]{fontenc}
\usepackage[colorlinks,urlcolor=blue,citecolor=blue,linkcolor=blue]{hyperref}

\newcommand{\vect}[1]{{\mathbf #1}}
\newcommand{\vegr}[1]{{\boldsymbol #1}}   
\newcommand{\kv}{{\bf k}}

\newcommand{\0}{{\bf 0}}
\renewcommand{\r}{{\bf r}}
\newcommand{\rhov}{{\bm \rho}}
\newcommand{\area}{\mathcal{A}}
\newcommand{\sch}{Schr{\"o}dinger }

\DeclareMathOperator*{\Sumt}{\tilde{\sum}}
\DeclareMathOperator*{\Sumb}{\Bar{\sum}}

\begin{document}

\title{Very strong light-matter coupling in patterned GaAs heterostructures}

\author{David de la Fuente Pico}
\thanks{D. F. P. and J. B. contributed equally to this work.}
\affiliation{Departamento de F\'isica Te\'orica de la Materia
  Condensada, Universidad
  Aut\'onoma de Madrid, Madrid 28049, Spain}
\affiliation{Condensed Matter Physics Center (IFIMAC), Universidad Autónoma de Madrid, 28049 Madrid, Spain}

\author{Johannes B\"urger}
\thanks{D. F. P. and J. B. contributed equally to this work.}
\affiliation{CNR Nanotec, Institute of Nanotechnology, via Monteroni, 73100 Lecce, Italy}

\author{Antonio Gianfrate}
\affiliation{CNR Nanotec, Institute of Nanotechnology, via Monteroni, 73100 Lecce, Italy}

\author{Jesper Levinsen}
\affiliation{School of Physics and Astronomy, Monash University, Victoria 3800, Australia}

\author{Meera M. Parish}
\affiliation{School of Physics and Astronomy, Monash University, Victoria 3800, Australia}

\author{Daniele Sanvitto}
\affiliation{CNR Nanotec, Institute of Nanotechnology, via Monteroni, 73100 Lecce, Italy}

\author{Francesca Maria Marchetti}
\affiliation{Departamento de F\'isica Te\'orica de la Materia
  Condensada, Universidad
  Aut\'onoma de Madrid, Madrid 28049, Spain}
\affiliation{Condensed Matter Physics Center (IFIMAC), Universidad Autónoma de Madrid, 28049 Madrid, Spain}

\author{Dario Ballarini}
\affiliation{CNR Nanotec, Institute of Nanotechnology, via Monteroni, 73100 Lecce, Italy}

\date{June 3, 2026}

\begin{abstract}
The very strong light–matter coupling regime enables the non-perturbative modification of matter properties via light. Using a patterned GaAs/AlGaAs waveguide with twelve wide quantum wells, we demonstrate hybridization of heavy- and light-hole excitons within a single polariton state and show that, at finite magnetic field, the presence of the light-hole exciton suppresses coupling to the heavy-hole Rydberg excitons and unbound scattering states.
We develop a fully microscopic theory that accounts for the combined effects of the magnetic field and light–matter coupling on the excitons, providing an accurate description of the experimental results beyond a perturbative coupled-oscillator framework.
This identifies 
quantum well width as a key control parameter for engineering the light-induced hybridization of matter wave functions in polaritons, which, in turn, can play a crucial role in the optical non-linearities.
\end{abstract}

\maketitle
The possibility of tailoring light by strongly coupling it to matter excitations is a central theme in semiconductor physics~\cite{Yamamoto_book2000}. The reversible energy transfer between photons confined in a cavity and excitons in a semiconductor leads to the formation of polariton quasiparticles with modified dispersion, enhanced nonlinearities, and efficient control~\cite{kavokin_2017microcavities}.
Strong light–matter coupling promotes the formation of condensates~\cite{Kasprzak-Marchetti_Nature2006,Deng-Haug-Yamamoto_RevModPhys2010}, collective coherence~\cite{Carusotto-RevMP2013}, and superfluid phases~\cite{Amo_Nature2009,Sanvitto_NatPhys2010}, as well as emergent technological applications such as low-threshold lasers and optical switching~\cite{Sanvitto-Kena-Cohen_NatMat2016,Schneider_NatComm2018}. These phenomena illustrate how the properties of light, such as its speed, coherence, and propagation, can be dramatically controlled by matter.

Equally important is the reverse approach of using light to modify the properties of matter, offering new perspectives to control and engineer quantum-matter phases~\cite{JBloch_Nature2022}.
In the context of polaritons, the very strong coupling regime~\cite{Khurgin_SSS2001}, where multiple matter excitations simultaneously hybridize with a single photonic mode, allows light to reshape the matter wave function itself~\cite{Zhang-Yamamoto_PRB2013,Yang_NJP2015,Brodbeck_PRL2017,Levinsen_PRR2019,Laird_PRB2022}. This effect expands the design space for polaritonic states, enabling control over excitonic properties, novel bound states~\cite{Cortese-DeLiberato_NatPhys2021,Kumar-Parish-Levinsen_PRB2022}, and manipulation of the spatial profile of sub-wavelength electromagnetic modes~\cite{Cortese_Optica2023}.
Thus far, the very strong coupling regime has been achieved in narrow ($7$ nm) GaAs multiple quantum wells (QWs) embedded in a microcavity~\cite{Brodbeck_PRL2017}. Using 
magneto-optical spectroscopy~\cite{Yang_NJP2015}, it was revealed that light significantly modified the exciton wave function in the upper polariton (UP)~\cite{Brodbeck_PRL2017}. 
However, in this case, hybridization was limited to exciton and continuum states of the same heavy-hole manifold~\cite{Laird_PRB2022}.
 
\begin{figure*}
\centering
\includegraphics[width=1.\linewidth]{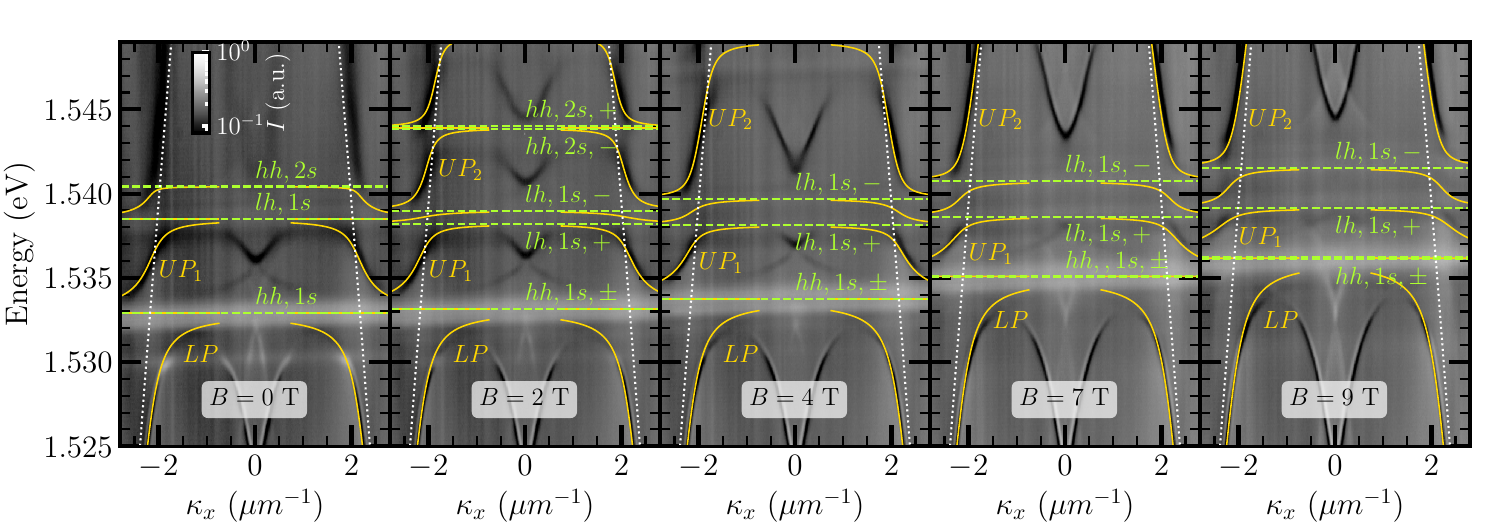}
    \caption{Reflectance spectra in the very strong coupling regime as a function of the photon in-plane wave vector $\kappa_x$ for increasing values of the magnetic field $B$. The polariton modes display anticrossing between the TE$_{1,\pm}$ waveguide modes (white dotted lines) and the the $hh$ and $lh$ exciton energies of states with different polarization, $E_{hh,ns,\pm}$ and  $E_{lh,1s,\pm}$ measured in weak coupling (horizontal dashed lines). 
    Solid (orange) lines are the theoretically predicted polariton energies.
    }
\label{fig:polariton-spectra-MF}
\end{figure*}

In this Letter, we use magneto-optical spectroscopy to probe the very strong coupling regime in patterned GaAs/AlGaAs waveguides, realizing a new form of light-induced matter hybridization involving both heavy- and light-hole excitons in wide QWs. 
Here, the $1s$ light-hole exciton is situated between the heavy-hole ground and excited states, such that the composition of the first UP is  fundamentally modified compared to the case of narrow QWs:
it is dominated by the $1s$ heavy-hole and $1s$ light-hole excitons, with a significant admixture of excited heavy-hole and continuum states that is reduced with increasing magnetic field.
This regime becomes accessible via a patterned-waveguide platform whose folded photonic dispersion enables simultaneous strong coupling to multiple excitonic Rydberg states (see Fig.~\ref{fig:polariton-spectra-MF}), a regime that is typically difficult to access in conventional planar microcavities~\cite{Rapaport_PRB2001,Pietka-Potemski_PRB2017}.
The interpretation of the experimental results are supported by quantitative agreement with a microscopic theory that non-perturbatively accounts for magnetic-field and light–matter coupling effects beyond a rigid coupled-oscillator description, validating this new form of light-induced matter hybridization.
These results establish a route to engineer hybridization between fundamentally different matter excitations,  thereby opening a new direction for light-engineered quantum matter.

%
\emph{Experimental setup.---}
The sample consists of a patterned GaAs/AlGaAs photonic waveguide heterostructure incorporating twelve GaAs QWs of width $d_z \simeq 20$~nm separated by Al$_{0.4}$Ga$_{0.6}$As barriers. Polarization-resolved white-light reflection measurements were performed with the sample mounted in a closed-cycle cryostat at $4$~K equipped with a superconducting magnet providing magnetic fields up to $B=9$~T in Faraday geometry. The structure was illuminated by a collimated thermal white-light source, and the reflected signal was collected by a low-temperature objective and directed to an imaging spectrometer, where the back focal plane was imaged to obtain momentum-resolved spectra. Polarization control was implemented either in excitation, to select TE and TM photonic modes, or in detection, using a circular polarization analyzer to resolve the Zeeman-split exciton transitions. Additional details are provided in the Supplemental Material (SM)~\cite{SM}.

\begin{figure}
    \centering
    \includegraphics[width=\columnwidth]{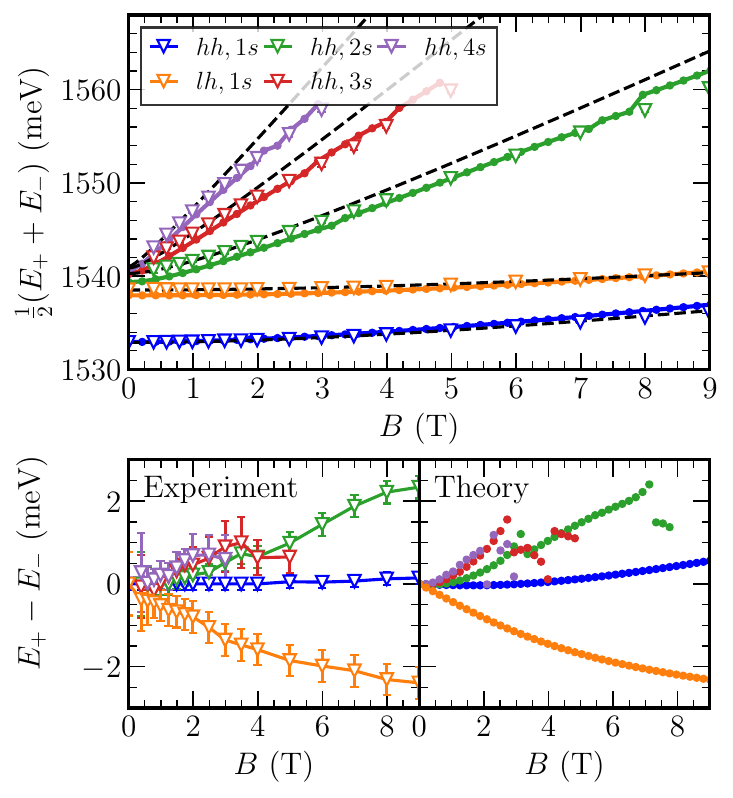}
    \caption{Top panel: Experimental exciton diamagnetic shift $(E_+ + E_-)/2$ as a function of magnetic field for the first five states (triangles). Points show theoretical results from the Luttinger Hamiltonian model without adjustable parameters (see text and Ref.~\cite{long-paper}), while dashed lines correspond to an effective 2D exciton model with fitted Rydberg energies and reduced masses (see~\cite{SM}). Bottom panels: Exciton Zeeman splitting $E_+ - E_-$ for the first five excited states.
    }
\label{fig:exciton-vs-MF}
\end{figure}
%
\emph{Model.---}
We consider wide GaAs QWs with heavy- ($hh$) and light-hole ($lh$) excitons that are strongly coupled to waveguide photons and subject to a perpendicular magnetic field. The microscopic Hamiltonian has three separate contributions,
$\hat{H} = \hat{H}_m + \hat{H}_{ph} + \hat{H}_{ph-m}$. 
The matter component $\hat{H}_m$,
\begin{multline}
\label{eq:matter}
    \hat{H}_m = \sum_{j,\bar{j}} \left[\int d\r \hat{\Psi}_j^\dag (\r)  {\mathcal{H}}_{j\bar{j}}^{} \hat{\Psi}_{\bar{j}}^{} (\r)\right.\\
    \left.+  \frac{1}{2} \int  d\r d\bar{\r} \hat{\Psi}_j^\dag (\r) \hat{\Psi}_{\bar{j}}^\dag (\bar{\r}) W_{j\bar{j}} (\r-\bar{\r}) \hat{\Psi}_{\bar{j}}^{} (\bar{\r}) \hat{\Psi}_j^{} (\r)\right]\, ,
\end{multline}
is expressed in terms of the conduction-electron and valence-hole creation and annihilation fermionic field operators, $\hat{\Psi}_j^{\dag} (\r)$ and $\hat{\Psi}_j^{} (\r)$---here, $\r=(x,y,z) = (\vegr{\rho},z) = (\rho,\theta,z)$ is the 3D coordinate and $\vegr{\rho}$ is the in-plane coordinate. The index $j=(J_e^{},J_h^{})$ labels conduction electrons and valence holes according to the $z$-component of their angular momentum: $J_e^{}=\pm 1/2$ for conduction electrons, and $J_h^{}=\pm 3/2$ and $\pm 1/2$ for heavy-hole ($hh$) and light-hole ($lh$) valence states, respectively~\cite{ivchenko2005optical,Haug_BookSemiconductors}.
The components of the single-particle Hamiltonian $ {\mathcal{H}}_{jj'}^{}$ for the electron part are diagonal (throughout we set $\hbar=1$),
\begin{equation}
     {\mathcal{H}}_{J_e^{} J_e^{}}^{} = E_g + 
    \frac{{ {\kv}_{e}}^2}{2m_e} + V_{e}(z) + J_e g_e \mu_B B\, ,
\end{equation}     
where $E_g$ is the bulk GaAs energy gap, $m_e$ the effective electron mass, $g_e$ is the electron $g$-factor, $\mu_B$ the Bohr magneton, and $B$ the strength of the perpendicular magnetic field. By contrast, the Luttinger Hamiltonian~\cite{Luttinger_PR1956}  describing the valence hole components,
\begin{align}
\label{eq:Luttinger}
     {\mathcal{H}}_{J_h^{} \bar{J}_h^{}}^{} &=  \begin{pmatrix}
        {h}_{\frac{3}{2}} &  {b} &  {c} & 0\\[5pt]
        {b}^{\,\dag} &  {h}_{\frac{1}{2}} & 0 &  {c}\\[5pt]
        {c}^{\,\dag} & 0 &  {h}_{-\frac{1}{2}} & - {b}\\[5pt]
       0 &  {c}^{\,\dag} & - {b}^{\,\dag} &  {h}_{-\frac{3}{2}}
   \end{pmatrix}_{J_h^{} \bar{J}_h^{}} \, ,
\end{align}
has both diagonal and off-diagonal terms:~\footnote{Note that in the term~\eqref{eq:c-term-lutt-ham} of the Luttinger Hamiltonian we have restored the in-plane isotropy, which is otherwise broken by the cubic crystal symmetry~\cite{Haug_BookSemiconductors}. This approximation significantly simplifies the numerical solution of the exciton problem, as discussed in Ref.~\cite{long-paper}.}
\begin{subequations}
\begin{align}
     {h}_{\pm\frac{3}{2}} &= \frac{ {k}_{z,h}^2}{2m_{z,hh}} + \frac{ {k}_{x,h}^2+ {k}_{y,h}^2}{2m_{hh}} + V_h (z) \mp 3\kappa \mu_B B\\
     {h}_{\pm\frac{1}{2}} &= \frac{ {k}_{z,h}^2}{2m_{z,lh}} + \frac{ {k}_{h,x}^2+ {k}_{h,y}^2}{2m_{lh}} + V_h (z)\mp \kappa \mu_B B\\
    \label{eq:b-term-lutt-ham}
     {b} &= -\frac{\sqrt{3} \gamma_3 }{2 m_0}\left\{ {k}_{h,z}, {k}_{h,x} -i {k}_{h,y}\right\}\\ 
    \label{eq:c-term-lutt-ham}
     {c} &= -\frac{\sqrt{3}}{2 m_0}\frac{\gamma_2 + \gamma_3}{2} \left( {k}_{h,x}- i {k}_{h,y}\right)^2 \, .
\end{align}
\end{subequations}
The $hh$ and $lh$ effective masses are, respectively,
\begin{align}
\label{eq:masses-gammas}
    m_{z,hh,lh} &= \frac{m_0}{\gamma_1 \mp 2 \gamma_2} & 
    m_{hh,lh} &= \frac{m_0}{\gamma_1 \pm \gamma_2} \, ,
\end{align}
where $m_0$ is the bare electron mass. The Luttinger parameters $\kappa$ and $\gamma_{i=1,2,3}$ are  
provided in the SM~\cite{SM}.
The potentials $V_{e,h}(z)$ describe the electron and hole QW confinements.

The momentum operators are
\begin{equation}
     {\kv}_{e,h} = -i \nabla_{\r} \pm \frac{e}{c} \vect{A}(\r)\, ,
\end{equation}
where, in the symmetric gauge, $\vect{A}(\r) = \frac{1}{2}\vect{B} \times \r$, with $\vect{B} = (0,0,B)$. This choice is  consistent with the Coulomb gauge, $\nabla \cdot \mathbf{A} = 0$, used for the light–matter coupling.
Finally, Coulomb interaction potentials are:
\begin{subequations}
\begin{align}
    W_{J_{e}^{} J_{h}^{}} (\r) &= -V(r) = -\frac{e^2}{\varepsilon r}\\
    W_{J_{e}^{} \bar{J}_{e}^{}} (\r) &= W_{J_{h}^{} \bar{J}_{h}^{}} (\r) = V(r)\, ,
\end{align}
\end{subequations}
where $r = \sqrt{\rho^2 + z^2}$, we use Gaussian units, $4\pi\epsilon_0 = 1$, and denote by $\varepsilon$ the static dielectric constant.

The term $\hat{H}_{ph}$ describes the TE waveguide modes, which are linearly polarized along the $y$-direction and propagate along $x$: 
\begin{equation}\label{eq:ham-ph}
    \hat{H}_{ph} = \omega \hat{a}^\dag \hat{a}^{}\, .
\end{equation}
The energy separation between TE and TM modes is sufficiently large ($\sim15$ meV) to allow us to neglect their coupling, resulting in the light–matter interaction:
\begin{equation}
\label{eq:ham-light-matter}
    \hat{H}_{ph-m} = \hat{a}^{\dag} \sum_{J_e, J_h} \frac{\lambda_{J_eJ_h}}{\sqrt{2 \area}}  \int d\r \hat{\Psi}_{J_e}^{} (\r) \hat{\Psi}_{J_h}^{} (\r) + \text{h.c.}\, ,
\end{equation}
where $\area$ is the in-plane system area, and the coupling strength satisfies the selection rules $\lambda_{J_e J_h} =\lambda_{\sigma}\delta_{\sigma}$, with
$\delta_{\sigma}=\delta_{\sigma,hh}\delta_{(J_{e}^{},J_{h}^{}) = (\mp1/2,\pm 3/2)}+\delta_{\sigma,lh}\delta_{(J_{e}^{},J_{h}^{}) = (\pm 1/2,\pm 1/2)}$. 
While angular-momentum algebra requires the light-matter couplings to be related via $\lambda_{lh} = \lambda_{hh}/\sqrt{3}$~\cite{ivchenko2005optical}, we treat the couplings independently and verify this relation \emph{a posteriori}.

Approximating the light-matter coupling 
as short ranged requires renormalizing the bare photon frequency $\omega$ to the dressed frequency $\nu$ 
and relating the couplings $\lambda_{\sigma}$ to the Rabi couplings $\Omega_{\sigma}$~\cite{Levinsen_PRR2019,Laird_PRB2022,deLaFuentePico_PRB2025,SM}. 
Furthermore, due to the large separation in momentum scale between waveguide photons and QW excitons, 
the dependence on the waveguide photon dispersion can be incorporated in the dressed photon frequency $\nu$.
We focus on the TE$_{1,\pm}$ modes, with energies $\nu = \nu_0 \pm v_g \kappa_x$ ($\nu_0 = 1619.5$~meV, $v_g = 61.31~\mu$m/ps), consistent with similar waveguides~\cite{Suarez-Forero2020, Ardizzone2022, Trypogeorgos_Nature2025}.  
Because the waveguide modes are linearly polarized, $\lambda_{\sigma}$ is independent of the electron–hole polarization $J_e + J_h=\pm1$. 

\emph{Excitons.---}
We first compare exciton magneto-optical measurements with theoretical predictions. 
We derive the exciton properties by solving the \sch equation $\hat{H}_{m} \hat{X}^{\dag}_{\alpha} \ket{0} = E \hat{X}^{\dag}_{\alpha} \ket{0} $ 
for the zero momentum exciton operator: 
\begin{multline}
\label{eq:exciton}
    \hat{X}^\dag_{\alpha=(J_z,J_e)} =\frac{1}{\sqrt{\area}} \Sumb_{J_h} \int d\r_e d\r_h e^{i\frac{e}{2c} (\vect{B} \times \vegr{\rho})\cdot \vect{R}_{J_h}} \\
    \times \varphi^{\alpha}_{J_h}(\vegr{\rho},z_e,z_h)\hat{\Psi}_{J_e}^\dag (\r_e) \hat{\Psi}_{J_h}^\dag (\r_h)\, ,
\end{multline}
where the exciton in-plane relative $\vegr{\rho} = \vegr{\rho}_e - \vegr{\rho}_h$ and center of mass $\vect{R}_{J_h} = (m_e \vegr{\rho}_e + m_{J_h} \vegr{\rho}_h)/(m_e + m_{J_h})$   coordinates  factorize thanks to the Lamb-Gor'kov-Dzyaloshinskii transformation~\cite{Lamb_PR1952,Gorkov_SovietJourn1968}---here, $m_{\pm 3/2} \equiv m_{hh}$ and $m_{\pm 1/2} \equiv m_{lh}$. As shown in Ref.~\cite{long-paper}, this approach, equivalent to that of Ref.~\cite{Bauer_AndoPRB1988}, accounts for exciton band mixing due to QW confinement and the magnetic field. 
Note that, while $J_h$ is not conserved due to $hh$–$lh$ mixing, in-plane isotropy preserves the angular-momentum projection $J_z = J_e + J_h + \ell$, where $\ell$ is the eigenvalue of $ \mathcal{L}_z = -i\partial_\theta$. This restricts hybridization (and the constraints in the sum, $\Sumb_{J_h}$) to 4-dimensional subspaces labeled by $\alpha = (J_z=\pm1, J_e=\pm1/2)$, where each component $J_h$ has a well defined value of $\ell$~\cite{SM}.

As shown in Fig.~\ref{fig:exciton-vs-MF}, the calculated exciton diamagnetic shift, $(E_+ + E_-)/2$, and Zeeman splitting, $E_+ - E_-$,  show excellent quantitative agreement with experiments for the first five states. Here, the polarization-resolved exciton energies $E_\pm$ for different magnetic fields~\footnote{We define the $+$ and $-$ polarizations following the theoretical convention, with the positive $z$ axis aligned along the growth direction (from the interior toward the top of the sample). Using the opposite convention as in, e.g., Ref.~\cite{Bauer-Ando2_PRB1988}, gives the same diamagnetic shift but the opposite Zeeman splitting.} are extracted from reflectance on a sample region without grating~\footnote{We select exciton states with the highest oscillator strength, thus relevant for polariton formation, by comparing reflectance spectra with and without the grating~\cite{SM}.}, corresponding to the weak-coupling regime. Notably, our microscopic model does not require any  adjustable parameters apart from a slightly reduced quantum-well width, $d_z = 18$~nm instead of the nominal $20$~nm, 
consistent with a slight off-center sample placement during growth, together with a rigid energy shift to align theory and experiment~\cite{SM}.
It can be shown~\cite{long-paper} that the five measured exciton states are primarily $s$-wave ($\ell = 0$) and mostly $hh$, except for the second state, which is mainly $lh$. Hybridization between different values of allowed quartets $(J_h,\ell)$~\cite{SM}  is strongest for the $lh$ state and grows for excited $hh$ Rydberg states at higher magnetic fields~\footnote{Experimental and theoretical results for the $hh, 3s$ and $hh, 4s$ states are limited to a restricted range of $B$ due to strong mixing with confined states of different symmetry (see Ref.~\cite{long-paper}).}.

These exact results are also compared with 
an effective 2D model, in which the $hh$ and $lh$ reduced masses and Rydberg energies serve as fitting parameters.
This approximation neglects the off-diagonal terms of the Luttinger Hamiltonian~\eqref{eq:Luttinger}, the $z$-dependence of the Coulomb interaction, and integrates out the $z$ degrees of freedom. In this limit, the exciton operator
$\hat{X}_\alpha^\dag \simeq \hat{X}_{\sigma,\pm}^\dag \delta_{J_e+J_h,\pm 1} \delta_{\sigma}$, 
depends only on the index $\sigma=hh,lh$ 
and on the light polarization $\pm$.
Within this simplified 2D description we obtain the Rydberg energies $R_{X\sigma}=2\mu_{\sigma}e^4/\varepsilon^2$ of $R_{Xhh}=8.45$~meV and $R_{Xlh}=10.8$~meV and  
the reduced masses, $\mu_\sigma=m_e m_\sigma/(m_e+m_\sigma)$, $\mu_{hh}=0.049\,m_0$ and $\mu_{lh}=0.063\,m_0$. The Rydberg energies exceed those of the full 3D model~\cite{long-paper} by $7.6\%$ and $13.8\%$, respectively. Discrepancies of the 2D model (Fig.~\ref{fig:exciton-vs-MF}) increase for higher Rydberg states and stronger magnetic fields due to enhanced mixing with dark excitons~\cite{long-paper}. 
Nevertheless, as discussed next, the very strong light–matter coupling induces $hh$–$lh$ hybridization that largely dominates over magnetic-field and confinement effects. We therefore employ the simpler 2D model in the polariton description below.

\emph{Polaritons.---} 
We finally examine waveguide magnetopolariton properties in the very strong coupling regime, and the possibility of hybridizing different exciton wave functions within the same polariton mode. 
The polariton states within the simplified 2D exciton model are defined as:
\begin{multline}
\label{eq:pol-state-main-text}
    |P^{} \rangle = \bigg[\sum_{\substack{\sigma,\pm}}\int d\rhov_e d\rhov_h e^{\frac{ie}{2c}(\mathbf{B}\times\rhov)\cdot \mathbf{R}_\sigma} \hat{\Psi}_{J_e}^{\dag}(\rhov_e)\hat{\Psi}_{J_h}^{\dag}(\rhov_h)\\\times\delta_\sigma\delta_{J_e+J_h,\pm1}\varphi_{\sigma,\pm}(\rho) +\gamma \hat a^{\dag} \bigg]\ket{0}\, .
\end{multline}
Polariton energies are found by minimizing $\bra{P}E - \hat{H} \ket{P}$ with respect to the electron-hole wavefunction $\varphi_{\sigma,\pm} (\rho)$ and the photon amplitude $\gamma$, where the wavefunction deviates from the bare exciton form of Eq.~\eqref{eq:exciton} and allows for light-induced modifications~\cite{Levinsen_PRR2019,Laird_PRB2022}. Here, the $z$ degrees of freedom have been integrated out of the Hamiltonian.

Polariton reflectance spectra are shown in Fig.~\ref{fig:polariton-spectra-MF} as a function of the waveguide wave vector and for increasing values of the magnetic field. Strong coupling is evidenced by the clear anticrossing pattern of the polariton modes as the TE${1,\pm}$ mode energies ($\nu$) cross the $hh$ and $lh$ exciton energies ($E_{hh,ns,\pm}$ and $E_{lh,1s,\pm}$).
Fitting the zero-magnetic-field polariton spectrum to a perturbative coupled oscillator model~\cite{SM} yields Rabi couplings of $\Omega_{hh}=5.41$~meV and $\Omega_{lh}=3.70$~meV for the $hh$ and $lh$ $1s$ excitons, respectively. These parameters constitute the only inputs to the 2D microscopic model. The large value $\Omega_{hh}\sim0.65R_{Xhh}$ places the system deep in the very strong light–matter coupling regime. In addition, the $hh$-$lh$ energy splitting $E_{lh,1s}-E_{hh,1s}\simeq5.44$~meV$\simeq\Omega_{hh}$ at $B=0$ and  decreases rapidly with $B$~\cite{SM}, causing light-induced hybridization of excitons to dominate over magnetic-field-induced hybridization.

\begin{figure}
    \centering
    \includegraphics[width=\columnwidth]{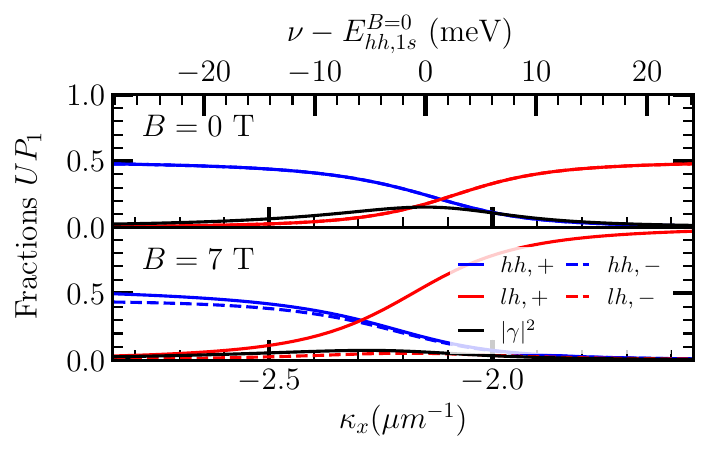}
    \caption{Exciton $|X_{\sigma,\pm}|^2=\langle 0 | \hat{X}_{\sigma,\pm} \hat{X}^{\dagger}_{\sigma,\pm}|0\rangle $ and photon $|\gamma|^2$ fractions~\eqref{eq:pol-state-main-text} of the $UP_1$ at $B=0,7$~T---different polarization components coincide at $B=0$.}
\label{fig:UP1-fractions_main-text}
\end{figure}

Because the $lh, 1s$ exciton state lies between the $hh$ Rydberg series, strong coupling with the waveguide mode implies that polariton modes contain a finite fraction of both $hh$ and $lh$ exciton components. 
We focus on the magnetic field evolution of the polariton state labeled as $UP_1$, whose energy interpolates between the exciton energies $E_{hh,1s,+} \simeq E_{hh,1s,-}$ and $E_{lh,1s,+}$. This state is mostly composed of $hh$ and $lh$ matter components, and its photon fraction, which is largest near resonance, decreases with increasing magnetic field (see Fig.~\ref{fig:UP1-fractions_main-text}).
We can quantify the degree of hybridization between exciton states by evaluating the overlap of the polariton state with the bare exciton states~\cite{SM}, concluding that, at finite magnetic field, light predominantly hybridizes the $hh,1s$ and $lh,1s$ exciton states in the $UP_1$ mode, whereas at zero magnetic field the hybridization also involves the $hh,2s$ state, higher Rydberg states, and the continuum of unbound states. This is a consequence of the small energy difference between the $lh,1s$ and the $hh,2s$ states at zero magnetic field, which is comparable to the coupling of the latter to the photon.
We conclude that the $lh,1s$ exciton, lying between the $hh,1s$ and $hh,2s$ states for wide QW~\cite{Rapaport_PRL2000,Rapaport_PRB2001,Cerda-Santos_PRApp2026}, strongly impacts the composition and properties of $UP_1$, unlike in narrow QWs where it lies well above the $hh$ Rydberg series~\cite{Laird_PRB2022}.

\begin{figure}
    \centering
    \includegraphics[width=\columnwidth]{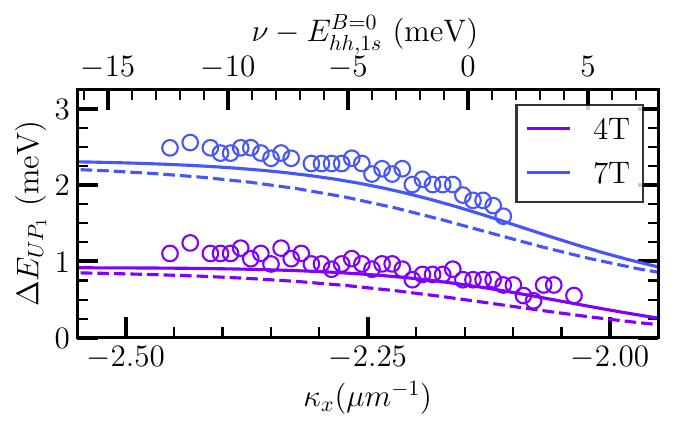}
    \caption{Comparison between the experimental (circles) and theoretical (solid lines) energy shift $\Delta E = E^{(B)} - E^{(B=0)}$ of the polariton state $UP_1$ at two magnetic fields for the TE$_{1,+}$ mode. Dashed lines show calculations including light coupling perturbatively (see text).}
\label{fig:UP-diamagnetic-shift}
\end{figure}

Signatures of wave function hybridization between $hh$ and $lh$ states can be quantified by measuring the polariton energy shift $\Delta E = E^{(B)} - E^{(B=0)}$ and comparing it with both our nonperturbative microscopic theory and a coupled-oscillator model, which treats the exciton as a point-like particle and thus only includes light–matter coupling perturbatively~\cite{SM}. 
Results are shown in Fig.~\ref{fig:UP-diamagnetic-shift}. Excellent agreement is found between experiment and theory, while the perturbative model underestimates the shift.

\emph{Discussion.---} 
We have employed magneto-optical spectroscopy to probe the very strong coupling between light and matter, revealing strong hybridization between heavy- and light-hole excitons and a photonic waveguide mode in patterned wide GaAs/AlGaAs QWs. Our microscopic theory quantitatively reproduces the experiments, confirming the very strong coupling regime.
This demonstrates that QW width provides a powerful knob for controlling light-induced matter hybridization between distinct electronic states.
An important perspective opened by our work is the exploration of the crucial role that hybridization can play in determining optical nonlinear properties.
Further, this platform enables the study of charged particles, such as trions~\cite{SM} arising from unintentional 
doping, and their coupling to light. Additionally, the application of a perpendicular electric field can create dipolar excitons with large dipole moments, opening the possibility to explore and control strongly interacting matter–light quasiparticles.

\begin{acknowledgments}
We would like to thank Paolo Cazzato for technical support with the experiments and the group of Loren N. Pfeiffer for the fabrication of the sample.  We are grateful to C. Tejedor and A. I. Fern\'andez-Dom\'inguez for fruitful discussions. DDFP acknowledges
financial support from the FPU Grant No.~FPU24/03084. DDFP and FMM acknowledge financial
support from the Spanish Ministry of Science, Innovation and
Universities through the ``Maria de Maetzu'' Programme for Units of
Excellence in R\&D (CEX2023-001316-M) and from the Comunidad de Madrid
and the Spanish State through the Recovery, Transformation and
Resilience Plan [``Materiales disruptivos bidimensionales 2D'' (MAD2D-
  CM)-UAM7], as well as from the Ministry of Science, Innovation and
Universities MCIN/AEI/10.13039/501100011033, FEDER UE, project
No.~PID2023-150420NB-C31 (Q). MMP is supported through Australian
Research Council Future Fellowship FT200100619, and JL through
Australian Research Council Discovery Project DP240100569. JB and DB acknowledge financial support under the National Recovery and Resilience Plan (NRRP) under the PRIN grant 2022, funded by the European Union– NextGenerationEU– Project Title PENNA- CUP B53D23003790006- Grant Assignment Decree No. 957 by the Italian Ministry of University and Research (MUR). JB, DB, and DS acknowledge financial support from “Quantum Optical Networks based on Exciton polaritons” (Q-ONE, N. 101115575, HORIZON- EIC- 2022- PATHFINDER CHALLENGES EU project), ”Neuromorphic Polariton Accelerator” (PolArt, N.101130304, Horizon-EIC-2023Pathfinder Open EU project), “National Quantum Science and Technology Institute” (NQSTI, N. PE0000023, PNRR MUR project), “Integrated Infrastructure Initiative in Photonic and Quantum Sciences” (I-PHOQS, N. IR0000016, PNRR MUR project). Views and opinions expressed are, however, those of the authors only and do not necessarily reflect those of the European Union or European Innovation Council and SMEs Executive Agency (EISMEA). Neither the European Union nor the granting authority can be held responsible for them.
\end{acknowledgments}

%

\newpage

\onecolumngrid
\clearpage
\begin{center}
\textbf{\large Supplemental Material: Very strong light-matter coupling in patterned GaAs heterostructures}\\
\vspace{4mm}
{David de la Fuente Pico,$^{1,2, *}$ Johannes B\"urger,$^{3,*}$  Antonio Gianfrate,$^{3}$ Jesper Levinsen,$^{4,5}$  Meera~M.~Parish,$^{4,5}$ Daniele Sanvitto,$^{3}$ Francesca Maria Marchetti,$^{1,2}$ and Dario Ballarini$^{3}$}\\
\vspace{2mm}
{\em \small
$^1$Departamento de F\'isica Te\'orica de la Materia
  Condensada, Universidad
  Aut\'onoma de Madrid, Madrid 28049, Spain \\
$^2$Condensed Matter Physics Center (IFIMAC), Universidad Autónoma de Madrid, 28049 Madrid, Spain\\
$^3$CNR Nanotec, Institute of Nanotechnology, via Monteroni, 73100 Lecce, Italy\\
$^4$School of Physics and Astronomy, Monash University, Victoria 3800, Australia\\
$^5$ARC Centre of Excellence in Future Low-Energy Electronics Technologies, Monash University, Victoria 3800, Australia\\
$^*$\rm{D. F. P. and J. B. contributed equally to this work
}}\end{center}
\setcounter{equation}{0}
\setcounter{figure}{0}
\setcounter{table}{0}
\setcounter{page}{1}
\makeatletter
\renewcommand{\theequation}{S\arabic{equation}}
\renewcommand{\thefigure}{S\arabic{figure}}


We present here the details of the sample and experimental setup used in the measurements, along with the procedure followed to extract the energies from the reflectance spectra. We outline the derivation of the exciton equations based on a Luttinger Hamiltonian, including a discussion of the choice of model parameters. We describe the simplification of the exciton description to an effective two-dimensional model and the procedure employed to extract the effective parameters by fitting the experimental data. This approach is then adopted to formulate the microscopic description of polaritons in the regime of very strong light–matter coupling. Finally, we provide additional information on the composition of the polaritons discussed in the main text.

\section{Sample}
\label{eq:sample}
\begin{figure}[h!]
\centering
\includegraphics[width=0.8\linewidth]{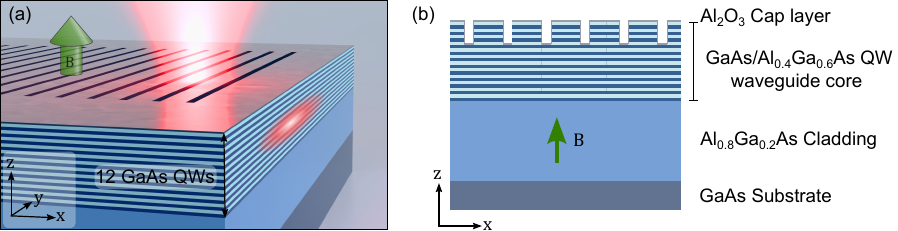}
    \caption{Geometry of the experimental sample. The waveguide core is formed by 12 $d_z \simeq 20$~nm wide GaAs/Al$_{0.4}$Ga$_{0.6}$As QWs separated by barriers. The etched subwavelength grating structure is along the $x$-direction (propagation direction of the guided mode). The magnetic field is applied in the $z$-direction and employed to tune and probe the exciton and polariton properties. (a)~3D view showing a simplified representation of the illumination scheme and waveguide mode. (b)~2D schematic side view depicting 6 unit cells of the waveguide. Dimensions are to scale, except for the substrate.}
\label{fig:exp_sample}
\end{figure}

%
The sample (schematically sketched in Fig.~\ref{fig:exp_sample}) is a GaAs/Al$_{0.4}$Ga$_{0.6}$As multiple-quantum-well photonic crystal waveguide grown by molecular beam epitaxy. The waveguide core consists of 12 GaAs quantum wells (QWs) ($d_z=20\,$nm thick) separated by Al$_{0.4}$Ga$_{0.6}$As barriers ($20\,$nm thick), arranged such that a barrier layer forms the top of the stack. The core is sandwiched between an Al$_{0.8}$Ga$_{0.2}$As lower cladding ($500\,$nm) grown on a GaAs substrate, and a $10\,$nm Al$_2$O$_3$ protective cap layer deposited on top.

A one-dimensional photonic crystal grating is defined over a $50 \times 400\,\mu$m$^2$ area by electron-beam lithography and dry etching. The grating has a period of $a = 242\,$nm, with a filling factor of approximately $75\%$, and an etch depth in the range $140$--$170\,$nm. The corresponding wave vector translation $\pi/a \approx 13\,\mu$m$^{-1}$ folds the guided modes back into the radiative light cone, enabling their observation in free-space reflection. At the wavelengths investigated here, both the fundamental TE$_0$ and the first-excited TE$_1$ waveguide modes are supported. We focus on the strong coupling regime between QW excitons and the TE$_1$ waveguide modes, as the resulting polariton modes provide stronger contrast in reflection measurements of the excited states. Details of the grating fabrication are given in Ref.~\cite{Riminucci_PRA2022, Forero2021}.

%

\begin{figure*}[ht]
\centering
\includegraphics[width=0.4\linewidth]{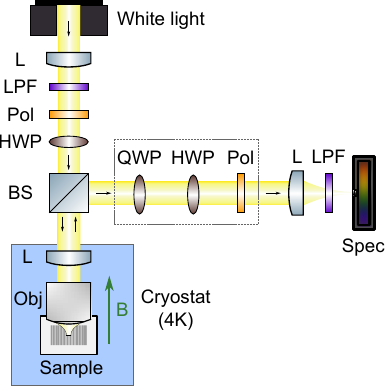}
	\caption{Optical setup for polarization-resolved white-light reflection measurements of magnetoexciton energies and magnetopolariton dispersions. The sample is mounted on a three-axis piezo translation stage inside a closed-cycle helium cryostat equipped with a superconducting magnet that generates a magnetic field parallel to the optical axis (i.e., perpendicular to the sample plane; indicated by the green arrow). Abbreviations: White light: thermal white-light source; L: lens; LPF: longpass filter; Pol: linear polarizer; HWP/QWP: half-/quarter-wave plate; BS: beamsplitter; Obj: objective; Spec: spectrometer. The dashed box indicates the circular polarization analyzer, which includes a HWP for compensating residual elliptical polarization. The circular analyzer was removed for polariton dispersion measurements. Optical component graphics are taken from~\cite{Franzen2006}.}
	\label{fig:exp_set-up}
\end{figure*}
%
\section{Experimental setup}
The optical setup used for polarization-resolved white-light reflection measurements of magnetoexciton energies and magnetopolariton dispersions is shown in Fig.~\ref{fig:exp_set-up}. The sample is mounted on a three-axis piezo translation stage inside a closed-cycle helium cryostat (attocube attoDRY~1000) operated at a base temperature of $4\,$K. The cryostat is equipped with a superconducting magnet capable of generating magnetic fields up to $9\,$T in Faraday configuration, i.e., with the field oriented parallel to the optical axis and perpendicular to the sample plane.

The sample is illuminated with a fiber-bundle-coupled tungsten-halogen white-light source (Thorlabs OSL2), collimated by a lens whose position is chosen to illuminate a large area of the waveguide grating structure. A low-temperature apochromatic objective (attocube LT-APO/633-RAMAN/0.81; clear aperture: $4.7\,$mm, focal length $f_{\mathrm{obj}} = 2.89\,$mm, numerical aperture: NA~$= 0.81$, working distance: $0.67\,$mm) focuses the illumination onto the sample and collects the reflected light, which is separated from the incident beam by a beamsplitter. The back focal plane of the objective is relayed onto the spectrometer entrance slit by a series of lenses with all but the last lens arranged in a 4$f$ configuration. The first relay lens ($f = 300\,$mm) is located inside the cryostat, followed by three external lenses ($f = 750\,$mm, $f = 300\,$mm, $f = 300\,$mm). This combination yields a total back focal plane magnification of $M_{\mathrm{lenses}} = 750/300 = 2.5$.

The reflected light is dispersed by an imaging spectrometer (Horiba iHR~550, focal length: $550\,$mm) equipped with a 600\,grooves/mm grating blazed at $500\,$nm and an entrance slit width of $94\,\mu$m. The dispersed signal is recorded by a 2D CCD camera (Hamamatsu ORCA-R2, pixel size: $6.45\,\mu\mathrm{m} \times 6.45\,\mu\mathrm{m}$) operated with $2\times2$ pixel binning and a typical exposure time of $105\,$s.
Furthermore, two $750\,$nm longpass filters are employed: one placed immediately after the light source to suppress the photoluminescence background from higher-energy excitation, and one immediately in front of the spectrometer entrance slit to suppress residual stray light. All measurements were performed in a darkened room.

Polarization control is achieved either in excitation, for measurements of the polariton dispersion or in detection, for measurements of exciton energies. For excitation-side control, a linear polarizer followed by a half-wave plate (HWP) is placed after the light source. Rotating the HWP selects either the TE or TM photonic modes of the waveguide.
For detection-side polarization analysis, a circular polarization analyzer is inserted into the reflected beam path after the beamsplitter. It consists of a quarter-wave plate (QWP) followed by a half-wave plate (HWP) and a linear polarizer (Pol). Rotating the HWP selects between the two circular polarization states ($+$ and $-$). 
The additional HWP serves to compensate for residual elliptical polarization introduced by the mirrors and other optical elements in the optical path.

\subsection{Momentum calibration}
The in-plane wave vector $\kappa_x$ is extracted from the CCD pixel position using
\begin{equation}
	\kappa_x = \frac{2\pi}{\lambda_0} \, \sin\!\left[\arctan\!\left(\frac{n_{\mathrm{px}} \, \Delta l_{\mathrm{px}} \, n_\mathrm{bin}}{M_{\mathrm{lenses}} \, M_{\mathrm{spec}} \, f_{\mathrm{obj}}}\right)\right],
	\label{eq:k_calibration}
\end{equation}
where $\lambda_0$ is the free-space wavelength, $n_{\mathrm{px}}$ is the pixel index on the CCD (with $n_{\mathrm{px}} = 0$ corresponding to $\kappa_x = 0$), $\Delta l_{\mathrm{px}} = 6.45\,\mu$m is the physical pixel size, $n_\mathrm{bin} = 2$ is the binning factor along the momentum axis, $M_{\mathrm{lenses}} = 2.5$ is the total magnification of the relay optics, $M_{\mathrm{spec}} = 1.1$ is the spectrometer magnification, and $f_{\mathrm{obj}} = 2.89\,$mm is the objective focal length. For these parameters, the argument of the trigonometric functions is small, so $\kappa_x$ is approximately proportional to $n_{\mathrm{px}}$. Nevertheless, the exact expression in Eq.~\eqref{eq:k_calibration} is used throughout all analyses.

\begin{figure*}
\centering
\includegraphics[width=0.7\linewidth]{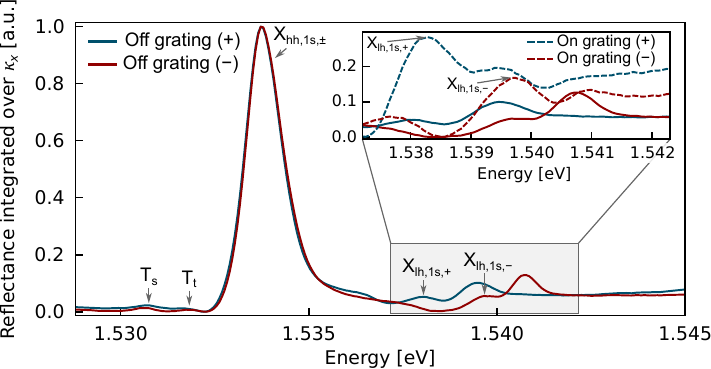}
	\caption{Reflectance spectrum at $B = 4$~T recorded on an unpatterned region of the sample for the two circular polarization states $+$ (blue) and $-$ (red). Substructure of the $lh, 1s$ peak is shown in the inset. For data recorded on the grating (dashed lines), the low-energy peak of the $lh$ doublet dominates. In contrast, on an unpatterned region of the sample (continuous lines), the high-energy peak dominates.
    }
\label{fig:reflectance-weak-coupling}
\end{figure*}
%
\subsection{Extraction of exciton and polariton energies}
Exciton energies are extracted from reflectance measurements performed on an unpatterned region of the sample ensuring the absence of strong light–matter coupling. In all exciton measurements, a circular polarization analyzer is used at the detection stage to independently acquire reflectance spectra for the $+$ and $-$ polarization states.
Spectra are collected at positions sufficiently far from the grating structure, thereby ensuring that the excitonic response is decoupled from any waveguide modes. To improve the signal to noise ratio, the reflectance data are integrated over the in plane wave vector $\kappa_x$, and the excitonic resonances are identified as spectral maxima.
For excitonic states that exhibit replica features, such as the $lh, 1s$ state which presents a characteristic double-peak structure, the extracted energies are compared with measurements performed in the grating region.
In this case, we also integrate over a finite range of $\kappa_x$ to improve the signal-to-noise ratio. Strictly speaking, the integration should be restricted to sufficiently large $\kappa_x$ values, corresponding to large exciton-waveguide detunings, where the system remains effectively in the weak-coupling regime. However, such a restriction substantially reduces the signal quality. We therefore extend the integration range while distinguishing excitonic and polaritonic features through their different signatures in the reflectance spectra: polaritonic modes appear as reflectance minima, whereas excitonic features manifest as maxima due to residual photoluminescence, as discussed below.
Since the contrast of the polariton dips is small near the exciton resonances, the integration effectively captures the exciton energies in the weak light–matter coupling regime as well-defined maxima.

\begin{figure}
    \centering
    \includegraphics[width=1\linewidth]{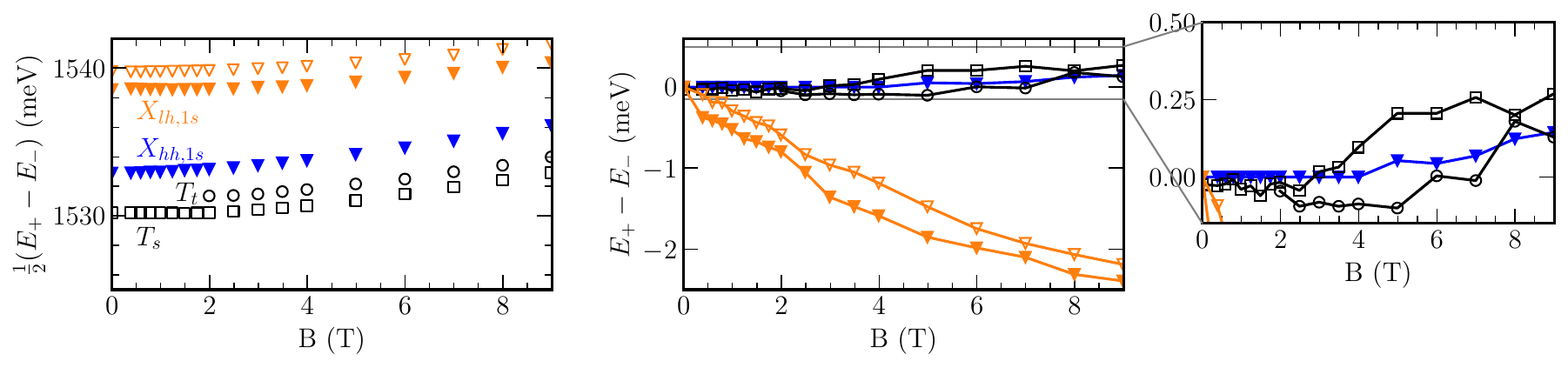}
    \caption{{Magnetic-field dependence of the experimentally measured diamagnetic shifts (left) and Zeeman splittings (right), extracted from the reflectance spectra in the weak-coupling regime. The filled triangles correspond to the $hh, 1s$ and $lh, 1s$ exciton states (same as those in Fig.~\ref{fig:exciton-vs-MF} of the main text), which correspond to the states with the highest oscillator strength. The empty (orange) triangles denote the $lh, 1s$ ``replica'' states with lower oscillator strength (see Fig.~\ref{fig:reflectance-weak-coupling}). The states with very small oscillator strength observed below the $hh, 1s$ state are compatible with singlet $T_s$ (black open squares) and triplet $T_t$ (black open circles) trion states.}}
\label{fig:additional-states-weak-coupling}
\end{figure}

A representative spectrum is shown in Fig.~\ref{fig:reflectance-weak-coupling}. We observe that, both on and off the grating, the dominant resonance corresponds to the $hh, 1s$ exciton, which exhibits the largest oscillator strength. At higher energies, the $lh, 1s$ exciton displays a doublet substructure, with the lower energy peak being more pronounced on the grating.
Because, on the grating, the lower energy component exhibits the larger oscillator strength, we use this component for comparison with theory.
By analyzing the magnetic field dependence of both the diamagnetic shift and the Zeeman splitting in Fig.~\ref{fig:additional-states-weak-coupling}, we find that the two components of the $lh, 1s$ doublet exhibit identical behavior. We attribute the observed doublet structure to optical interference effects in the multilayer stack, where multiple reflections may give rise to interference fringes.
We rule out alternative explanations based on disorder or variations in the width of the 12 QWs. While the energy separation between the $hh, 1s$ state and the lower energy component of the doublet is consistent with a confinement width of $d_z = 18$~nm, the higher energy component would require an unrealistically small width of approximately $15$~nm.

Below the $hh, 1s$ exciton, the reflectance spectra in Fig.~\ref{fig:reflectance-weak-coupling} exhibit very weak spectral features. Their magnetic field dependence is shown in Fig.~\ref{fig:additional-states-weak-coupling}. These features are consistent with singlet ($T_s$) and triplet ($T_t$) trion states. Indeed, both positively and negatively charged singlet trions ($T_s$), as well as triplet trions ($T_t$), have been reported to exhibit larger Zeeman splittings while maintaining diamagnetic shifts comparable to those of the neutral exciton in doped GaAs QWs~\cite{Vanhoucke_PRB2002,Shields_PRB1995,Peeters_PhysB2001}.
While the presence of these states is not relevant for the analysis of the $UP$ branch discussed in the main text (see Fig.~\ref{fig:LP-anticrossing_checkUP} and the discussion therein), they give rise to weak anticrossings with the waveguide modes and modify the dispersion of the lower polariton ($LP$) branches, as analyzed below.

\begin{figure}
    \centering
    \includegraphics[width=\linewidth]{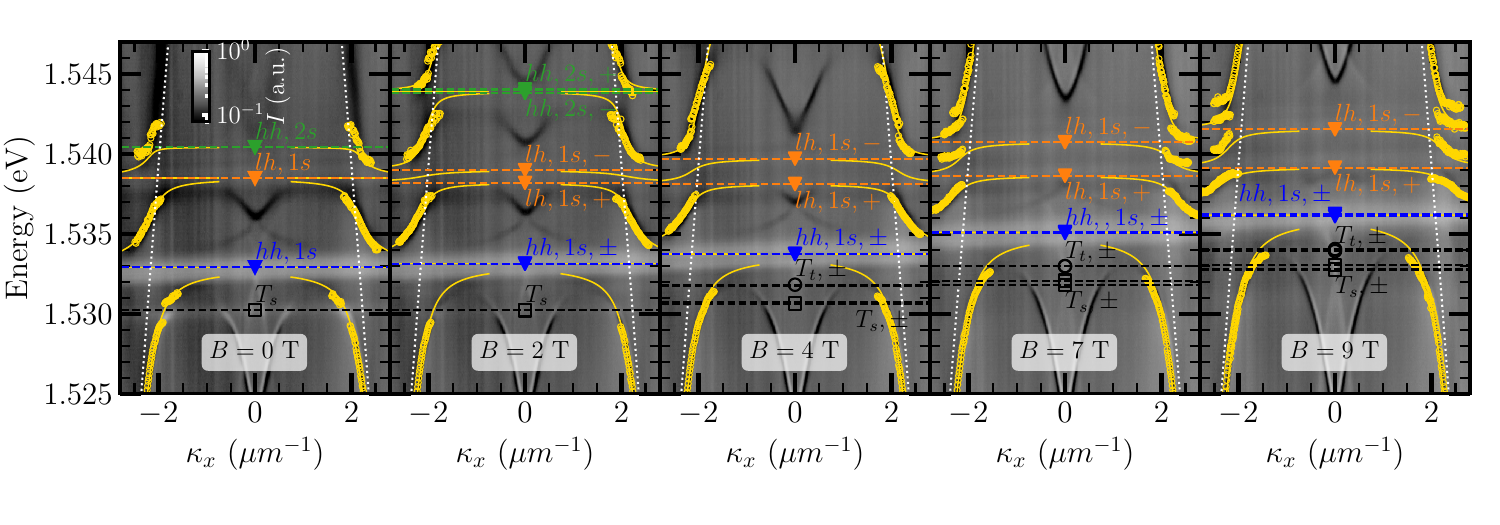}
    \caption{Polariton reflectance spectra for increasing values of the magnetic field $B$. Circles (yellow) are the energies of the polariton modes extracted by evaluating the reflectance spectrum second derivative with respect to energy. Solid (yellow) lines indicate the theoretically predicted polariton energies {obtained from the microscopic 2D polariton model described below in this Supplementary Material and in the main text}. Horizontal lines are experimental exciton and trion energies from Fig.~\ref{fig:additional-states-weak-coupling} and Fig.~\ref{fig:exciton-vs-MF} of the main text.}
\label{fig:additional-states-strong-coupling}
\end{figure}

Polariton dispersions are measured on the grating. We focus on the strong coupling regime between matter excitations with TE waveguide modes, which are selected by the linear polarizer in the excitation path.
The polariton branches appear as dips in the angle-resolved reflectance spectra. To reliably identify the weak dips associated with excited-state polariton branches, dip positions are extracted as the maxima of the second derivative of the reflectance with respect to energy. The second derivative suppresses the slowly varying spectral background and enhances the weaker excited-state dips relative to the dominant lower-polariton feature, enabling consistent extraction of the dispersion for all polariton branches using the same procedure. 
Fig.~\ref{fig:additional-states-strong-coupling} shows the reflectance spectrum colormap (the same as in Fig.~\ref{fig:polariton-spectra-MF} of the main text) with the polariton energies extracted from the second derivative superimposed. In addition, the exciton energies, as well as the singlet and triplet trion energies measured under weak coupling, are included.
We observe that, for the $lh, 1s$ exciton doublet seen in weak coupling (Fig.~\ref{fig:reflectance-weak-coupling}), only the lower-energy component couples strongly to light. This is expected, as this component has the largest oscillator strength on the grating under weak-coupling conditions.
In addition, we observe anticrossing features of the $LP$ with the singlet trion ($T_s$) at low magnetic fields and with the triplet trion ($T_t$) at higher magnetic fields. 
The anticrossing with $T_s$ disappears as the field increases, whereas the more pronounced bending of the $LP$ at high fields likely arises from the anticrossing with $T_t$.
Our work focuses on the very strong coupling properties of exciton states, in particular on the first upper polariton state, labeled as $UP_1$ in the main text. 
In the later section of this SM, \emph{Trion states and lower polariton dispersion}, we show that the weak anticrossing features of the $LP$ do not affect the energy of the upper polariton states. This observation justifies the use of our polariton model, which excludes a microscopic treatment of trion states.


\section{Luttinger model for excitons in a magnetic field}
\label{sec:exciton-Luttinger}
In this section, we derive the \sch equation describing excitons in a GaAs/Al$_{0.4}$Ga$_{0.6}$As quantum well (QW) of width $d_z$ subject to a perpendicular magnetic field $\vect{B} = (0,0,B)$, which includes the effects of hybridization between bands due to confinement as well as magnetic field effects. To this end we employ the matter Hamiltonian $\hat{H}_m$~\eqref{eq:matter}, which includes the Luttinger Hamiltonian for the valence holes~\eqref{eq:Luttinger}~\cite{Haug_BookSemiconductors,ivchenko2005optical}, as well as the ansatz~\eqref{eq:exciton} for the exciton with zero magnetic momentum.

We start from the matter Hamiltonian~\eqref{eq:matter} of the main text and the ansatz~\eqref{eq:exciton} for the exciton states $\hat{X}^{\dag}_{\alpha} \ket{0}$, satisfying the normalisation condition:
\begin{equation}
    1= \bra{0} \hat{X}^{}_{\alpha} \hat{X}^{\dag}_{\alpha} \ket{0} = \Sumb_{J_h} \int dz_e dz_h d\vegr{\rho} \left|\varphi_{J_h}^\alpha (\vegr{\rho},z_e,z_h)\right|^2 \, .
\end{equation}
The eigenvalue problem $\hat{H}_m\hat{X}^{\dagger}\ket{0} = E \hat{X}^{\dagger}\ket{0}$ can be equivalently written as a \sch equation for the four components of the exciton wave function in Eq.~\eqref{eq:exciton}, $\varphi^{\alpha}_{J_h}(\vegr{\rho},z_e,z_h)$:
\begin{equation}
\label{eq:Luttinger-exciton}
    \begin{pmatrix}
         {h}_{X\frac{3}{2}} &  {b}_X &  {c}_X & 0\\[5pt]
        {b}^{\,\dag}_X &  {h}_{X\frac{1}{2}} & 0 &  {c}_X\\[5pt]
        {c}^{\,\dag}_X & 0 &  {h}_{X-\frac{1}{2}} & - {b}_X\\[5pt]
       0 &  {c}^{\,\dag}_X & - {b}^{\,\dag}_X &  {h}_{X-\frac{3}{2}}
   \end{pmatrix} \begin{pmatrix}
       \varphi^{\alpha}_{\frac{3}{2}}(\rhov,z_e,z_h) \\[5pt] \varphi^{\alpha}_{\frac{1}{2}} (\rhov,z_e,z_h) \\[5pt] \varphi^{\alpha}_{-\frac{1}{2}} (\rhov,z_e,z_h) \\[5pt] \varphi^{\alpha}_{-\frac{3}{2}} (\rhov,z_e,z_h)
   \end{pmatrix} -\frac{e^2}{\varepsilon \sqrt{\rho^2 + (z_e-z_h)^2}} \begin{pmatrix}
       \varphi^{\alpha}_{\frac{3}{2}}(\rhov,z_e,z_h) \\[5pt] \varphi^{\alpha}_{\frac{1}{2}} (\rhov,z_e,z_h) \\[5pt] \varphi^{\alpha}_{-\frac{1}{2}} (\rhov,z_e,z_h) \\[5pt] \varphi^{\alpha}_{-\frac{3}{2}} (\rhov,z_e,z_h)
   \end{pmatrix}
   = E \begin{pmatrix}
       \varphi^{\alpha}_{\frac{3}{2}}(\rhov,z_e,z_h) \\[5pt] \varphi^{\alpha}_{\frac{1}{2}} (\rhov,z_e,z_h) \\[5pt] \varphi^{\alpha}_{-\frac{1}{2}} (\rhov,z_e,z_h) \\[5pt] \varphi^{\alpha}_{-\frac{3}{2}} (\rhov,z_e,z_h)
   \end{pmatrix}\, , 
\end{equation}
where $\alpha=(J_z,J_e)$ denotes the set of conserved quantum numbers: These are the $z$-projection of the total angular momentum $J_z$, which we define below, and the conduction electron angular momentum $J_e=\pm 1/2$. Coupling to light limits the values of $J_z$ to  $J_z=\pm 1$. Instead, the hole angular momentum,
$J_h=\pm 3/2, \pm 1,2$, is not a good quantum number, as we explain below. The exciton Luttinger Hamiltonian matrix elements are given by~\cite{Bauer_AndoPRB1988}:
\begin{subequations}
\begin{align}
     {h}_{X\pm\frac{3}{2}}= &-\frac{\partial_{z_h}^2}{2m_{z,hh}}-\frac{\partial_{z_e}^2}{2m_e}-\frac{1}{2\mu_{hh} } \left(\partial_\rho^2+
    \frac{\partial_\rho}{\rho}+\frac{\partial_\theta^2}{\rho^2}\right) 
    + \frac{1}{2} \mu_{hh} \omega_{chh}^2 
    \rho^2 + V_e(z_e) + V_h(z_h) \nonumber\\
    & -\frac{eB}{2\mu_{hh}c}\eta_{hh} i\partial_\theta\mp3\kappa\mu_B B + g_e J_e \mu_B B \\
     {h}_{X\pm\frac{1}{2}} = &-\frac{\partial_{z_h}^2}{2m_{z,lh}}-\frac{\partial_{z_e}^2}{2m_e}-\frac{1}{2\mu_{lh} } \left(\partial_\rho^2+
    \frac{\partial_\rho}{\rho}+\frac{\partial_\theta^2}{\rho^2}\right)
    + \frac{1}{2} \mu_{lh} \omega_{clh}^2 
    \rho^2 + V_e(z_e) + V_h(z_h) \nonumber\\
    & -\frac{eB}{2\mu_{lh}c}\eta_{lh} i\partial_\theta\mp\kappa\mu_B B + g_eJ_e\mu_B B \\ 
     {b}_X = &  i\frac{\sqrt{3}\gamma_3}{m_0} {k}^{-}\partial_{z_h}\\
     {c}_X = & -\frac{\sqrt{3}}{2 m_0}
    \frac{\gamma_2+\gamma_3}{2}( {k}^{-})^2
    \, ,
\end{align}
\end{subequations}
where 
\begin{subequations}
\begin{align}
     {k}^{\pm} &= i e^{\pm i\theta}\left(\partial_\rho\pm\frac{i}{\rho}\partial_\theta\pm\frac{eB}{2c}\rho\right) \\
    ( {k}^{\pm})^2 &= e^{\pm i2\theta}\left[-\frac{e^2B^2}{4c^2}\rho^2+\frac{eB}{c}\left(\mp\rho\partial_\rho-i\partial_\theta \right) - \left( \partial_\rho^2\pm\frac{2i}{\rho}\partial_\theta\partial_\rho\mp\frac{2i\partial_{\theta}}{\rho^2}-\frac{\partial_\rho}{\rho}-\frac{\partial_\theta^2}{\rho^2} \right)\right]\, .
\end{align}
\end{subequations}
The expressions of the $hh$ and $lh$ masses, $m_{z,\sigma}$ and $m_{\sigma}$ in terms of the Luttinger parameters $\gamma_{1,2}$ are given in Eq.~\eqref{eq:masses-gammas} of the main text, while the exciton reduced masses and mass differences are defined as
\begin{align}
    \mu_{\sigma} &= \frac{m_e m_\sigma}{m_e + m_\sigma} & \eta_{\sigma} & =\frac{m_e-m_\sigma}{m_e+m_\sigma}\, ,
\end{align}
where $\sigma=hh,lh$. 
The exciton cyclotron frequencies are defined as
\begin{equation}
\label{eq:cyclotron}
    \omega_{c\sigma} = \frac{eB}{2\mu_\sigma c}\, .
\end{equation}
The explicit expressions of $ {b}^{\,\dag}_{X}, {c}^{\,\dag}_{X} $ are immediate since $( {k}^{-})^{\dag}= {k}^{+}$.

The QW confining potentials for electrons and holes are assumed to exhibit sharp boundaries and are given by:
\begin{equation}
    V_{e,h} (z) = V_{c,v}\theta\Big(\frac{d_z}{2}-z\Big)\, ,
\end{equation}
where $\theta(x)$ is the Heaviside theta function and $V_{c,v}$ are the valence and conduction band offsets. 
The band offsets, $V_c$ and $V_v$, represent the fractions of the bandgap difference $\Delta E_g$ between the well material (GaAs) and the barrier material 
(Al$_{0.4}$Ga$_{0.6}$As) that are assigned to the conduction and valence bands, so that $\Delta E_g=V_c+V_v$. 
In our case, we consider that $\Delta E_g$ is partitioned between the conduction-band offset $V_c$ 
and the valence-band offset $V_v$ according to the  ratio $V_c:V_v=67:33$~\cite{Bataev-Ignatiev-Efimov_PRB2022}. Further, we neglect the dielectric-constant mismatch and mass discontinuity between the well and the barrier. For a GaAs/Al$_{0.4}$Ga$_{0.6}$As QW of width 
$d_z=18$~nm, taking into account the dielectric-constant mismatch leads to a binding-energy corrections of the order of $0.4$~meV, while including mass discontinuity effects leads to a correction of the order of $0.1$~meV~\cite{Andreani-Pasquarello_PRB1990,Shuvayev_2006}. The full set of parameters employed is summarized in Tab.~\ref{tab:parameters-model}. 
Note that we have selected all the parameters listed in Tab.~\ref{tab:parameters-model}, except for the well width $d_z$ from the literature. These are standard parameters typically used to describe GaAs/Al$_{0.4}$Ga$_{0.6}$As QWs~\cite{Bauer_AndoPRB1988,Bataev-Ignatiev-Efimov_PRB2022,Vurgaftman-GaAsPar_JAP2001}. As discussed in detail towards the end of this section, we fix the well width to $d_z=18$~nm rather than to the value specified by the grower, $d_z=20$~nm, because it optimally describes the experimental exciton energies. 
Finally, note that we have already simplified the more general expression of the term
\begin{equation*}
     {c}_X = -\frac{\sqrt{3}}{4 m_0}\left[(\gamma_2-\gamma_3)( {k}^{+})^2+ (\gamma_2+\gamma_3)( {k}^{-})^2\right] \, ,
\end{equation*}
so that to neglect the in-plane cubic anisotropy~\cite{Bastard_1999JQE,Bauer_AndoPRB1988}. 
In this case, assuming in-plane isotropy simplifies the problem considerably, as explained next.

\begin{table}[h!]
\centering
\begin{tabular}{l l c}
\hline
 & \textbf{Parameter} &  \textbf{Value} \\
\hline\hline
electron mass 
 & $m_{e}$ ($m_0$) & $0.067$ \\
dielectric constant
 & $\varepsilon$ & 12.5 \\
electron $g$-factor
 & $g_e$ & -0.44 \\
 \hline
\multirow{4}{*}{Luttinger parameters} 
 & $\gamma_{1}$ & 6.98 \\
 & $\gamma_{2}$ & 2.06 \\
 & $\gamma_{3}$ & 2.93 \\
 & $\kappa$ & 1.2 \\
\hline
\multirow{3}{*}{Band offset} 
 & $\Delta E_g$~(eV)  & 0.5  \\
 & $V_v (\Delta E_g)$ & $0.33$ \\
 & $V_c  (\Delta E_g)$ & $0.67$ \\
\hline
Well width
 & $d_z$ (nm) & 18\\
\hline
\end{tabular}
\caption{Choice of parameter values specific to the GaAs/Al$_{0.4}$Ga$_{0.6}$As QW employed in experiments that have been used to solve the exciton \sch equation in either real~\eqref{eq:Luttinger-exciton}~ or reciprocal space~\eqref{eq:Luttinger-exciton-momentum}. None of these parameters, apart from the effective well width, were chosen to better describe the experiments. Rather, these are standard parameters already employed to describe GaAs/Al$_{0.4}$Ga$_{0.6}$As QWs. In particular,  the Luttinger parameters, the electron mass, and the electron $g$-factor have been taken from Ref.~\cite{Vurgaftman-GaAsPar_JAP2001}  and Ref.~\cite{Bauer_AndoPRB1988}.
At $4$~K, the values of the bandgaps of bulk GaAs and Al$_{0.4}$Ga$_{0.6}$As are $E_g^{\text{bulk}}(\text{GaAs})=1.519$~eV and $E_g^{\text{bulk}}(\text{Al$_{0.4}$Ga$_{0.6}$As})\simeq2.055$~eV~\cite{Vurgaftman-GaAsPar_JAP2001}. This gives a bandgap difference
$\Delta E_g\simeq0.5$~eV. The bandgap discontinuity is assumed to be partitioned between the conduction and valence bands in a $67:33$ ratio, following Ref.~\cite{Bataev-Ignatiev-Efimov_PRB2022}, so that $V_v+ V_c = \Delta E_g$.}
\label{tab:parameters-model}
\end{table}

In fact, although the $z$-component of the orbital angular momentum, $\mathcal{L}_z = -i\partial_\theta$, does not commute with the $ {b}_X$ and $ {c}_X$ terms---so its eigenvalue $\ell$ is not a good quantum number---the in-plane isotropy ensures that the $z$-component of the total exciton angular momentum is conserved.
Specifically, $J_z = J_h + J_e + \ell$ remains a good quantum number, as explained in the main text. Because the electron component $J_e = \pm 1/2$ is also conserved, 
the in-plane isotropy reduces the exciton Hilbert space to 4-dimensional subspaces with a well defined value of the index $\alpha=(J_z,J_e)$---see Tab.~\ref{tab:ell-Jh-quartets}. Note that, within each allowed quartet, only the single $s$-wave $(J_h,\ell=0)$ combination has a finite oscillator strength. Hybridization with this component, allows the dark states to gain a finite coupling to light. 
\begin{table}[h]
\centering
\renewcommand{\arraystretch}{1.15}
\setlength{\tabcolsep}{6pt}
\begin{tabular}{c cc||cccc c}
\hline
$\alpha =$ & $J_z$ & $J_e$ 
& $\tfrac{3}{2}$ & $\tfrac{1}{2}$ 
& $-\tfrac{1}{2}$ & $-\tfrac{3}{2}$ 
& $=J_h$ \\
\hline\hline
&  1  & $-\tfrac{1}{2}$ &  0  &  1  &  2  &  3  & \multirow{4}{*}{$=\ell$} \\
&  1  & $+\tfrac{1}{2}$ & -1  &  0  &  1  &  2  & \\
& -1  & $-\tfrac{1}{2}$ & -2  & -1  &  0  &  1  & \\
& -1  & $+\tfrac{1}{2}$ & -3  & -2  & -1  &  0  & \\
\hline
\end{tabular}
\caption{Quartets of allowed values of the pairs $(J_h,\ell)$ which can hybridize within each $\alpha=(J_z,J_e)$ sector of good quantum numbers---here $(J_e^{},J_h^{})$ are the $z$-component of the angular momentum of conduction electrons and valence holes, respectively, while $J_z = J_h + J_e + \ell$ is  the $z$-component of the total exciton angular momentum, which is conserved.}
\label{tab:ell-Jh-quartets}
\end{table}

Due to the long-range nature of the Coulomb interaction, we find it more convenient to numerically solve the exciton-coupled \sch equations~\eqref{eq:Luttinger-exciton} in the in-plane reciprocal space, rather than in real space. To achieve this, we developed an efficient subtraction trick~\cite{Laird_PRB2022,deLaFuentePico_PRB2025} that significantly improves convergence. We have also generalized this method for the case of a finite QW in the accompanying paper~\cite{long-paper}. To this end, we rewrite~\eqref{eq:Luttinger-exciton} by considering the Fourier transform
\begin{equation}
    \varphi^{\alpha}_{J_h} (\vegr{\rho},z_e,z_h) = \frac{1}{\sqrt{\area}} \sum_{\kv} \varphi^{\alpha}_{J_h \kv} (z_e,z_h) e^{i\kv\cdot \vegr{\rho}}\, ,
\end{equation}
thus obtaining:
\begin{multline}
\label{eq:Luttinger-exciton-momentum}
    \begin{pmatrix}
         {h}_{X\frac{3}{2}} &  {b}_X &  {c}_X & 0\\[5pt]
        {b}^{\,\dag}_X &  {h}_{X\frac{1}{2}} & 0 &  {c}_X\\[5pt]
        {c}^{\,\dag}_X & 0 &  {h}_{X-\frac{1}{2}} & - {b}_X\\[5pt]
       0 &  {c}^{\,\dag}_X & - {b}^{\,\dag}_X &  {h}_{X-\frac{3}{2}}
   \end{pmatrix} \begin{pmatrix}
       \varphi^{\alpha}_{\frac{3}{2}\kv}(z_e,z_h) \\[5pt] \varphi^{\alpha}_{\frac{1}{2}\kv}(z_e,z_h) \\[5pt] \varphi^{\alpha}_{-\frac{1}{2}\kv}(z_e,z_h) \\[5pt]\varphi^{\alpha}_{-\frac{3}{2}\kv}(z_e,z_h)
   \end{pmatrix} + \sum_{\kv'} V_{\kv-\kv'} (z_e-z_h) \begin{pmatrix}
       \varphi^{\alpha}_{\frac{3}{2}\kv}(z_e,z_h) \\[5pt] \varphi^{\alpha}_{\frac{1}{2}\kv}(z_e,z_h) \\[5pt] \varphi^{\alpha}_{-\frac{1}{2}\kv}(z_e,z_h) \\[5pt]\varphi^{\alpha}_{-\frac{3}{2}\kv}(z_e,z_h)
   \end{pmatrix}\\
   = E \begin{pmatrix}
       \varphi^{\alpha}_{\frac{3}{2}\kv}(z_e,z_h) \\[5pt] \varphi^{\alpha}_{\frac{1}{2}\kv}(z_e,z_h) \\[5pt] \varphi^{\alpha}_{-\frac{1}{2},\kv}(z_e,z_h) \\[5pt]\varphi^{\alpha}_{-\frac{3}{2}\kv}(z_e,z_h)
   \end{pmatrix}\, ,    
\end{multline}
where we employ the same notation as in Eq.~\eqref{eq:Luttinger-exciton} but  the terms of the Hamiltonian are now given in momentum space:
\begin{subequations}
\label{eq:Ham-Luttinger-exciton-momentum}
\begin{align}
     {h}_{X\pm\frac{3}{2}}= &-\frac{\partial_{z_h}^2}{2m_{hh,z}}-\frac{\partial_{z_e}^2}{2m_e} + \frac{k^2}{2\mu_{hh} } 
    + \frac{1}{2}\mu_{hh} \omega_{c hh}^2 
    \left(-\partial_k^2 - \frac{\partial_k}{k} - \frac{\partial_\theta^2}{k^2}\right) + V_e(z_e) + V_h(z_h) \nonumber\\
    & -\frac{eB}{2\mu_{hh}c}\eta_{hh} i\partial_\theta\mp3\kappa\mu_B B + g_e J_e \mu_B B \\
     {h}_{X\pm\frac{1}{2}} = &-\frac{\partial_{z_h}^2}{2m_{lh,z}}-\frac{\partial_{z_e}^2}{2m_e} + \frac{k^2}{2\mu_{lh} }
    + \frac{1}{2}\mu_{lh} \omega_{c lh}^2 
    \left(-\partial_k^2 - \frac{\partial_k}{k} - \frac{\partial_\theta^2}{k^2}\right) + V_e(z_e) + V_h(z_h) \nonumber\\
    & -\frac{eB}{2\mu_{lh}c}\eta_{lh} i\partial_\theta\mp\kappa\mu_B B + g_eJ_e\mu_B B \\ 
     {b}_X&=i\frac{\sqrt{3}\gamma_{3}}{m_0}\partial_{zh}  {k}^{-}\\
\label{eq:c-coupling-lutt-k}    
     {c}_X&= - \frac{\sqrt{3}}{2m_0}  \frac{\gamma_2+\gamma_3}{2}( {k}^{-})^2\,,
\end{align}
\end{subequations}
and
\begin{align}
     {k}^{\pm}=& e^{\pm i\theta}\Big(-k+\frac{eB}{2c}(\mp\partial_k-i\frac{\partial_\theta}{k})\Big)\\
    ( {k}^{\pm})^2=& e^{\pm 2i\theta} \Bigg[k^2 - \frac{eB}{2c} \left(\mp2k \partial_k -2i \partial_\theta\right) + \left(\frac{eB}{2c}\right)^2 \left(\partial_k^2 \mp \frac{2i}{k^2} \partial_\theta \pm\frac{2i}{k} \partial_{k\theta} - \frac{\partial_k}{k} - \frac{\partial_\theta^2}{k^2}\right)\Bigg].
\end{align}
Again, the complex conjugate terms of the $ {b}_X$ and  $ {c}_X$ are readily obtained using $ {k}^{+}=( {k}^{-})^{\dagger}$. The normalisation condition for $\varphi_{Jh\kv}^\alpha (z_e,z_h)$ now reads as
\begin{equation}
    1= \Sumb_{J_h} \int dz_e dz_h \sum_{\kv} \left|\varphi_{J_h\kv}^\alpha (z_e,z_h) \right|^2\, .
\end{equation}

In order to find the four coupled equations to solve within the $4$-dimensional space of wave functions with definite values of $\alpha = (J_z, J_e)$, we expand the exciton wave function on the basis of orbital angular momentum $\ell$,
\begin{equation}
    \varphi^{\alpha}_{J_h\kv} (z_e,z_h) = 
    e^{i\ell \theta} \varphi^{\alpha}_{J_h k \ell} (z_e,z_h)\delta_{\ell , J_z -J_e- J_h}\, ,
\end{equation}
and project each component onto the state with definite value of $\ell$ by multiplying Eq.~\eqref{eq:Luttinger-exciton} by $\int_0^{2\pi} \tfrac{d\theta}{2\pi} e^{-i\ell' \theta}$. The explicit form of these equations, as well as the numerical method employed for its solution can be found in the accompanying paper~\cite{long-paper}.

The set of microscopic model parameters employed to find the exciton energies is summarized in Tab.\ref{tab:parameters-model}. 
As discussed in the main text, the quantitative agreement of our theoretical results for the exciton
Zeeman splitting is excellent for the first five exciton states. It is important to stress that the agreement is obtained without any adjustable parameters. Only the QW width has been chosen to a slightly smaller value than the nominal one specified by the growers, i.e., we fix $d_z=18$~nm, while the expected width is $d_z=20$~nm. 
According to the growers’ indications, a deviation of $2$~nm is reasonable, as the layer thickness at the center of the deposition region is typically slightly higher than at the edges; the effective QW thickness thus depends on the sample’s position during deposition.
Note that, despite the omission of additional effects in the sample, such as disorder, mechanical strain at the interfaces, doping, and defects, the choice of parameters in Tab.~\ref{tab:parameters-model} allows to reproduce the experimental results remarkably well.
The quantitative agreement of our results for the exciton diamagnetic shift is also remarkably good, aside from an overall shift of the energy scale. In particular, using the well-known value for bulk GaAs, $E_g(\text{GaAs}) = 1519$~meV at $T = 4$~K, our theory predicts the $hh, 1s$ exciton energy to be $E_{hh,1s}^{B=0} = 1527.2$~meV for a well width $d_z = 18$~nm. The experimental value, however, is $E_{hh,1s}^{B=0, \text{exp}} = 1532.9$~meV, resulting in a difference of $5.7$~meV. This discrepancy could be attributed to a rigid shift of the exciton lines introduced by additional doping. Indeed, as discussed previously, below the $hh, 1s$ exciton energy we observe additional states compatible with singlet and triplet charged trion states, $T_{s,t}$.

\section{Fitting the exciton diamagnetic shift with a simplified 2D exciton model}
\label{sec:exciton-2D}
We describe in this section a simplified purely two-dimensional (2D) model that characterizes magnetoexcitons and employs the exciton reduced mass and binding energy as fitting parameters in order to reproduce the experimental diamagnetic shift of both ground and excited exciton states. This model has been successfully employed to describe excitons in narrow GaAs QWs (see Ref.~\cite{Laird_PRB2022}). For wider QWs, as in our experimental setup, we obtain excellent agreement for the $hh$ and $lh$ exciton ground states. However, the agreement progressively worsens for higher Rydberg states and at stronger magnetic fields because of the enhanced mixing described in the previous section. 

The 2D description of excitons in a magnetic field consists in neglecting the off-diagonal terms of the Luttinger Hamiltonian in the \sch equation~\eqref{eq:Luttinger-exciton}, neglecting the $z$-dependence of the Coulomb interaction and integrating out the $z$ degrees of freedom. In this limit, the $z$-component of the exciton orbital angular momentum, $ {\ell}_z$, is again conserved, so that only the s-wave $\ell=0$ exciton states couple to light. As a consequence, $J_h$ is also a good quantum number, and we can classify excitons according to the value of $J_h$: states with $J_h = \pm 3/2$ correspond to $\sigma=hh$ excitons, while those with $J_h = \pm 1/2$ correspond to $\sigma=lh$ excitons. They are further characterized by the light polarization $J_z = \pm 1$ to which they couple: 
\begin{equation}
    \varphi_{J_h}^{\alpha=(J_z,J_e)} (\vegr{\rho},z_e,z_h) = \delta_{J_z,J_e+J_h}\delta_{J_z, \pm 1} \delta_\sigma \varphi_{\sigma} (\rho) \xi_{e}(z_e) \xi_{h}(z_h)\, ,
\end{equation}
where
\begin{equation}
\label{eq:delta-sigma}
    \delta_\sigma = \begin{cases}
        \delta_{(J_{e}^{},J_{h}^{}) = (\mp\frac{1}{2},\pm \frac{3}{2})} & \sigma=hh\\
        \delta_{(J_{e}^{},J_{h}^{}) = (\pm\frac{1}{2},\pm \frac{1}{2})} & \sigma=lh\, ,
    \end{cases}
\end{equation}
and where $\xi_{e,h}(z_{e,h})$ are the wave functions describing the QW confinement of electrons and holes---$\int dz |\xi_{e,h}(z)|^2 = 1$.
We thus write the exciton wave function as $\varphi_{\sigma}(\rho)$, where $\sigma=(hh,lh)$. 
The \sch equation~\eqref{eq:Luttinger-exciton} that describes them now simplifies to
\begin{equation}
\label{eq:2D-exciton}
    \left[E_{g\sigma} -\frac{1}{2\mu_{\sigma}} \left(\frac{d^2}{d\rho^2}+
    \frac{1}{\rho}\frac{d}{d\rho}\right)
    + \frac{1}{2} \mu_{\sigma}  \omega_{c\sigma}^2  \rho^2 -\frac{e^2}{\varepsilon \rho} \pm g_{\sigma} \mu_B B\right] \varphi_{\sigma} (\rho) = E_{\pm} \varphi_{\sigma} (\rho)\, ,  
\end{equation}
where $E_{g\sigma}$ are the effective $hh$ and $lh$ energy gaps which include the effects of confinement once we have intergrated out the $z$ degrees of freedom---note that these completely factorise if we neglect the off-diagonal terms in the Luttinger Hamiltonian, as well as the $z$-dependence in the Coulomb interaction:
\begin{equation*}
    E_{g\sigma} = \int dz_e dz_h \xi_e^* (z_e) \xi_h^* (z_h) \left[-\frac{\partial_{z_e}^2}{2m_{\sigma,z}} - \frac{\partial_{z_e}^2}{2 m_e} + V_e(z_e) + V_h (z_h) \right] \xi_e^{} (z_e) \xi_h^{} (z_h)\, .
\end{equation*}
The exciton wave functions satisfy the normalization condition:
\begin{equation}
    1= \int d\vegr{\rho} \left|\varphi_{\sigma} (\rho)\right|^2\, .
\end{equation}
The $\sigma = hh, lh$ exciton cyclotron frequencies $\omega_{c\sigma}$ are defined in Eq.~\eqref{eq:cyclotron} and $g_{\sigma}$ are the effective Zeeman $g$-factors. 
According to Eq.~\eqref{eq:Ham-Luttinger-exciton-momentum}, we have that $g_{hh}=-3\kappa-g_e/2$ and $g_{lh} = -\kappa + g_e/2$ and, by using the parameters of Tab.~\ref{tab:parameters-model}, $g_{hh} \simeq -3.38$ and $g_{lh} \simeq -1.42$, which do not correspond to the values measured in experiments---see  Fig.~\ref{fig:exciton-vs-MF} of the main text. This is because, in order to correctly describe the exciton Zeeman splitting and their different values for different Rydberg states, it is necessary to include the off-diagonal terms of the Luttinger Hamiltonian, how demonstrated by the very good agreement shown with theory in Fig.~\ref{fig:exciton-vs-MF}. Note also that, because the Zeeman term is described as a rigid shift in the 2D exciton model~\eqref{eq:2D-exciton}, the exciton wave function $\varphi_{\sigma}(\rho)$ does not depend on the light polarization index $\pm$.
The Zeeman terms do not contribute to the exciton diamagnetic shift, which is what we use to compare theory with experiments to extract the effective parameters of Eq.~\eqref{eq:2D-exciton}.

An efficient method to numerically solve Eq.~\eqref{eq:2D-exciton} has been already presented in Ref.~\cite{Laird_PRB2022}. Here, we follow the same procedure so that to extract the exciton parameters by fitting the experimental exciton diamagnetic shift, as described next. 
The energy scales characterizing the $hh$ and $lh$ exciton problem are those related to Coulomb interaction, thus the Rydberg energies
\begin{equation}
\label{eq:hydrogen_scales}
    R_{X\sigma} =\frac{2\mu_{\sigma}  e^4}{\varepsilon^2}=\frac{1}{2\mu_{\sigma}  a_{X\sigma}^2}\, , 
\end{equation}
where $a_{X\sigma} = \varepsilon/(2\mu_\sigma e^2)$ are the two Bohr radii, 
and to the magnetic field, i.e., the cyclotron frequencies $\omega_{c\sigma}$. Note that the $hh$ and $lh$ Rydberg energies need to satisfy the constraint
\begin{equation}
\label{eq:constraints-hh-lh-rydbergs}
    \frac{R_{Xlh}}{R_{Xhh}} = \frac{\mu_{lh}}{\mu_{hh}} \, .
\end{equation}
The reduced masses $\mu_\sigma$ can be fitted by the magnetic field dependence of the  diamagnetic shifts of the $1s$ $hh$ and $lh$ states, $E_{hh,1s}^{(B)}$ and $E_{lh,1s}^{(B)}$. For the $lh$ exciton, the fit is restricted to the range $B>2$~T, where the experimental resolution is higher. The value of the Rydberg energy $R_{Xhh}$ has been extracted by comparing the experimental and theoretical value of the energy difference $E_{hh,2s}^{(B=0.4~\text{T})} - E_{hh,1s}^{(B=0.4~\text{T})}$. The value of $R_{Xlh}$ is instead derived from the constraint~\eqref{eq:constraints-hh-lh-rydbergs}. Finally, the effective energy gaps $E_{g\sigma}$ are extracted so that to fix the values of $E_{hh, 1s}^{(B=0)}, E_{lh, 1s}^{(B=2~\text{T})}$ to the experimental ones. The extracted parameters are listed in Tab.~\ref{tab:hh-lh_summary}. 
\begin{table}[h]
    \centering
    \begin{tabular}{l c c}
    \hline
    \textbf{Parameter} & \multicolumn{2}{c}{\textbf{Value}} \\ 
     & $hh$ & $lh$ \\ 
    \hline \hline
    Exciton Rydberg energy $R_{X\sigma}$ (meV) & 8.45 & 10.8 \\  
    Exciton reduced mass $\mu_{\sigma}~(m_0)$ & 0.049 & 0.063 \\ 
    Energy gap $E_{g\sigma}$ (eV) & 1.541 & 1.549 \\  \hline
    Rabi coupling $\Omega_{\sigma}$ (meV) & 5.41 & 3.70 \\ \hline
\end{tabular}
    \caption{Parameters for the $hh$ and $lh$ excitons obtained by fitting the experimental diamagnetic shift data with an effective 2D model. In the last row of the table we report the values of the Rabi $\Omega_{\sigma}$ obtained by fitting the polariton energies with a coupled oscillator model~\eqref{eq:5-coupled oscillator model}, as explained in the next section of this SM.}
\label{tab:hh-lh_summary}
\end{table}
%

\section{Modeling 2D polaritons in the very strong coupling regime}
\label{sec:2D-polaritons}
In this section we introduce the effective 2D model that we employ to describe the very strong light-matter coupling regime and polaritons. 
As discussed in the main text, the Hamiltonian for heavy- and light-hole excitons in a GaAs QW subject to a magnetic field applied perpendicular to the well plane can be decomposed into three contributions: the matter term, $\hat{H}_m$, describing the electron and hole degrees of freedom; the photonic term $\hat{H}_{ph}$, accounting for the TE waveguide mode; and the light-matter coupling term between an electron-hole pair and the guided mode $\hat{H}_{ph-m}$~\eqref{eq:ham-light-matter} :
\begin{align}
    \hat{H} &= \hat{H}_m + \hat{H}_{ph} + \hat{H}_{ph-m}\, .
\end{align}
In 2D, the analogous of the 3D matter Hamiltonian~\eqref{eq:matter} considered in the main text reads as:
\begin{equation}
\label{eq:2Dmatter}
    \hat{H}_m = \sum_{j,\bar{j}} \left[\int d\vegr{\rho} \hat{\Psi}_j^\dag (\vegr{\rho}) \frac{\left[-i\nabla \pm  \frac{e}{c} \vect{A} (\vegr{\rho})\right]^2}{2m_j}\hat{\Psi}_{\bar{j}}^{} (\vegr{\rho}) +  \frac{1}{2} \int  d\vegr{\rho} d\bar{\vegr{\rho}} \hat{\Psi}_j^\dag (\vegr{\rho}) \hat{\Psi}_{\bar{j}}^\dag (\bar{\vegr{\rho}}) W_{j\bar{j}} (\vegr{\rho}-\bar{\vegr{\rho}}) \hat{\Psi}_{\bar{j}}^{} (\bar{\vegr{\rho}}) \hat{\Psi}_j^{} (\vegr{\rho})\right]\, ,
\end{equation}
where $\hat{\Psi}_j$ ($\hat{\Psi}^{\dag}_j$) are the fermionic annihilation (creation) field operators in 2D, satisfying the anticommutation relations $\{\hat{\Psi}_j^{} (\rhov), \hat{\Psi}_{\bar{j}}^{\dag} (\bar{\rhov})\}=\delta_{j,\bar{j}} \delta(\rhov - \bar{\rhov})$ and $\{\hat{\Psi}_j^{} (\rhov), \hat{\Psi}_{\bar{j}}^{} (\bar{\rhov})\}=0$, where $j=(J_e^{},J_h^{})$ as before. 
The Coulomb interaction terms are now purely 2D:
\begin{subequations}
\begin{align}
    W_{J_{e}^{} J_{h}^{}} (\rhov) &= -V(\rho) = -\frac{e^2}{\varepsilon \rho}\\
    W_{J_{e}^{} \bar{J}_{e}^{}} (\rhov) &= W_{J_{h}^{} \bar{J}_{h}^{}} (\rhov) = V(\rho)\, .
\end{align}
\end{subequations}
The photonic term is the same as in the maint text~\eqref{eq:ham-ph},
\begin{equation}  
    \hat{H}_{C}= \omega \hat{a}^\dag \hat{a}^{}\, ,
\end{equation}
where $\hat{a}$ ($\hat{a}^{\dag}$) is the annihilation  (creation) operator of a single mode photon. 
As explained in the main text, we effectively incorporate the dispersion of the waveguide mode through the photon frequency $\omega$. Finally, the light-matter Hamiltonian in 2D reads as
\begin{equation}
\label{eq:2Dham-light-matter}
    \hat{H}_{ph-m} = \hat{a}^{\dag} \Sumt_{J_e, J_h} \frac{\lambda_{J_eJ_h}}{\sqrt{2 \area}}  \int d\r \hat{\Psi}_{J_e}^{} (\rhov) \hat{\Psi}_{J_h}^{} (\rhov) + \text{h.c.} \, ,
\end{equation}
where $\area$ is the in-plane system area, and  $\lambda_{J_eJ_h}$ is the coupling strength defined in Eq.~\eqref{eq:ham-light-matter}, which obeys the selection rules 
$\Sumt = \sum \delta_{J_{e}^{} + J_{h}^{}, \pm 1}$, where $\pm$ corresponds to light polarization:
\begin{equation}\label{eq:g-coupling-def}
   \lambda_{J_eJ_h} = \begin{cases}
       \lambda_{hh} & (J_{e}^{},J_{h}^{}) = (\mp\frac{1}{2},\pm \frac{3}{2})\\
       \lambda_{lh} & (J_{e}^{},J_{h}^{}) = (\pm\frac{1}{2},\pm \frac{1}{2})\, .
   \end{cases} 
\end{equation}

With these approximations, the polariton state is written as:
\begin{equation}
\label{eq:pol-state}
    |P^{} \rangle = \left[ \sum_{\substack{\sigma,\pm}}\int d\rhov_e d\rhov_h e^{\frac{ie}{2c}(\mathbf{B}\times\rhov)\cdot \mathbf{R}_\sigma} \hat{\Psi}_{J_e}^{\dag}(\rhov_e)\hat{\Psi}_{J_h}^{\dag}(\rhov_h)\delta_\sigma\delta_{J_e+J_h,\pm1}\varphi_{\sigma,\pm}(\rho) +\gamma \hat a^{\dag} \right]\ket{0}\, .
\end{equation}
where $\delta_\sigma$ is defined in Eq.~\eqref{eq:delta-sigma}.
Normalization $\braket{P}=1$ requires that
\begin{equation}
\label{eq:normalization-P}
    \sum_{\sigma,\pm}\int d\vegr{\rho}  |\varphi_{\sigma,\pm} (\rho)|^2 +|\gamma|^2=1.
\end{equation}

The polariton coupled equations can be obtained by minimizing the functional $\bra{P}E - \hat{H} \ket{P}$, giving:
\begin{subequations}
\label{eq:sigma-polariton-eqs}
\begin{align}
\label{eq:pol1}
     E \varphi_{\sigma,\pm}(\rho) &= \left[E_{\sigma} - \frac{1}{2\mu_{\sigma} } \left(\frac{d^2}{d\rho^2}+
    \frac{1}{\rho}\frac{d}{d\rho}\right)
    + \frac{1}{2} \mu_{\sigma}  \omega_{c \sigma}^2 \rho^2 - \frac{e^2}{\varepsilon \rho} \pm g_{\sigma}\mu_B B\right]\varphi_{\sigma,\pm}(\rho) 
     + \frac{\lambda_{\sigma}}{\sqrt{2}} \gamma \delta(\rhov)\\
\label{eq:pol2}
    E\gamma &= \omega\gamma  + \sum_{\sigma,\pm} \frac{\lambda_\sigma}{\sqrt{2}} \int d\rhov \varphi_{\sigma,\pm}(\rho) \delta(\rhov)\, .
\end{align}
\end{subequations}
{Note that, while for the purely excitonic equation~\eqref{eq:2D-exciton} the wave function $\varphi_{\sigma}(\rho)$ does not depend explicitly on the light polarization $\pm$, in the polariton case the coupling to light restores this dependence in the electron-hole pair wave function describing the polariton state, $\varphi_{\sigma,\pm}(\rho)$.} 
The presence of the Dirac delta $\delta(\rhov)$ in Eq.~\eqref{eq:pol1} introduces a logarithmic divergence in the excitons wave functions $\varphi_{\sigma,\pm}(\rho) \sim \log(\rho)$ as $\rho \to 0$. This ultraviolet divergence requires a renormalization of the bare photon waveguide energy $\omega$, as explained in detail in Ref.~\cite{Levinsen_PRR2019} in absence of a magnetic field. Note that because the divergence
occurs at zero electron-hole separation, $\vegr{\rho} = \0$, where the magnetic field has no effect, the renormalisation procedure follow the same steps of Ref.~\cite{Levinsen_PRR2019}---for the case of magnetopolaritons in narrow GaAs QW structures see Ref.~\cite{Laird_PRB2022}, while Ref.~\cite{deLaFuentePico_PRB2025} considers the case of TMD monolayers. We generalize the scheme here to the case of 4 exciton states (two $\sigma=hh,lh$ modes and two $\pm$ polarization modes) coupled to a single  photonic waveguide mode.

The divergent part of $\phi_{\sigma,\pm}(\rho)$ can be rescaled out of the exciton wave function:
\begin{equation}
\label{eq:rescale}
    \varphi_{\sigma,\pm}(\rho) = \beta_{\sigma,\pm}(\rho) - \frac{\lambda_\sigma \mu_{\sigma} \gamma}{\sqrt{2}\pi}K_0\left(\frac{\rho}{a_{X\sigma}}\right)\, ,
\end{equation}
where $K_0(x)$ is the zeroth-order modified Bessel function of the second kind. One can then show that the divergence is then renormalized by redefinyning the bare photon energy $\omega$ in terms of the experimentally measurable dressed energy $\nu$
\begin{equation}
\label{eq:def-delta-real-space}
    \nu =\omega - \sum_{\sigma, \pm}\frac{\lambda_\sigma^2 \mu_{\sigma}}{2 \pi}\int d\rhov K_0\left(\frac{\rho}{a_{X\sigma}}\right)\delta(\rhov) \, .
\end{equation}
It is the renormalized photon energy that is related to the dispersion of the two TE modes that strongly coupled to the exciton modes:
\begin{equation}
    \nu = \nu_0 \pm v_g \kappa_x\, ,
\end{equation}
where $\nu_0=1619.5$~meV, $v_g=61.31$~$\mu$m/ps.
Further, the light-matter coupling strengths $\lambda_\sigma$ can be related to the Rabi couplings $\Omega_\sigma$ via~\cite{Levinsen_PRR2019}
\begin{equation}
\label{eq:Rabi-c}
    \Omega_{\sigma} = \lambda_{\sigma} \varphi_{\sigma, 1s}^{(B=0)}(0) = \frac{\lambda_{\sigma}}{a_{X\sigma}}\sqrt{\frac{2}{\pi}}\, ,
\end{equation}
where $\varphi_{\sigma, 1s}^{(B=0)}(0)$ are the $hh$ and $lh$ hydrogenic exciton wave functions in absence of a magnetic field evaluated a zero relative electron and hole distance. 

The coupled polariton equations~\eqref{eq:sigma-polariton-eqs} together with the renormalization scheme, i.e., by using Eqs.~\eqref{eq:def-delta-real-space} and~\eqref{eq:Rabi-c}, become: 
\begin{subequations}
\label{eq:renorm-polaritons}
\begin{align}
\label{eq:r-pol1}
    E \varphi_{\sigma,\pm}(\rho) &= \left[E_{\sigma} - \frac{1}{2\mu_{\sigma} } \left(\frac{d^2}{d\rho^2}+
    \frac{1}{\rho}\frac{d}{d\rho}\right)
    + \frac{1}{2} \mu_{\sigma}  \omega_{c \sigma}^2 \rho^2 - \frac{e^2}{\varepsilon \rho} \pm g_{\sigma}\mu_B  B \right]\varphi_{\sigma,\pm}(\rho)  + \frac{\sqrt{\pi}\Omega_{\sigma} a_{X\sigma}}{2} \gamma \delta(\rhov)\\
    E\gamma &= \nu\gamma  + \sum_{\sigma,\pm} \left[ \frac{\sqrt{\pi}\Omega_\sigma a_{X\sigma}}{2} \int d\rhov \varphi_{\sigma,\pm}(\rho) \delta(\rhov) + \frac{\pi\Omega_{\sigma}^2}{4R_{X\sigma}} \int d\rhov K_0 \left(\frac{\rho}{a_{X\sigma}}\right) \delta(\rhov)\gamma\right]\, .
\end{align}
\end{subequations}
The above system of five coupled differential equations allow for a miscroscopic description of polaritons well into the very strong coupling regime, where the exciton component of each polariton state can hybridize between several Rydberg exciton states, as well as the continuum. The equations are 
solved using the same numerical procedure already detailed Ref.~\cite{Laird_PRB2022} and the results are those employed to derive the theoretical polariton energies of Fig.~\ref{fig:polariton-spectra-MF} in the main text, as well as the upper polariton diamagnetic shift of Fig.~\ref{fig:UP-diamagnetic-shift}.
We use the parameters listed in Tab.~\ref{tab:hh-lh_summary}.
We adopt the computational approach introduced in Ref.~\cite{Laird_PRB2022}. This technique takes advantage of a mapping between the 2D harmonic oscillator and the 2D hydrogen-like system, enabling an efficient numerical treatment of both the full Rydberg spectrum and the excitonic ground state. In our case is generalized to four types of exciton modes, $hh$ and $lh$, and with two possible polarizations, $\pm$. 

The parameters listed in Tab.~\ref{tab:hh-lh_summary} are those necessary for a correct quantitative description of the $hh, 1s$, $lh, 2s$ exciton states, as well as the $hh, 2s$ state at moderate magnetic fields. 
For the Zeeman terms, $g_{\sigma} \mu_B B$, we use $g_{\sigma}$ values extracted from experimental measurements of the $1s, hh$ and $1s, lh$ states.
In practice, this is equivalent to using an effective magnetic-field-dependent $g_{\sigma}(B)$. As discussed in the previous section of this SM, this simplified model assigns the same Zeeman splitting to all excited exciton Rydberg states with the same $\sigma$ character. This constitutes a limitation of the present approach for a quantitative description of upper polariton states that interpolate with the $hh, 3s$ and higher excited states. Indeed, the Zeeman splitting of the $hh, 2s$ state is small ($\lesssim 1$~meV). Since the main focus of our results is the first excited upper polariton state, which interpolates between the $hh, 1s$ and $lh, 1s$ states, our approach still provides quantitatively accurate results, as demonstrated by the excellent agreement between the experimental and theoretical dispersions of all polariton states shown in Fig.~\ref{fig:polariton-spectra-MF} of the main text.
We now describe the procedure used to extract the values of $\Omega_{hh,lh}$ in Tab.~\ref{tab:hh-lh_summary} from experimental measurements.

An important consideration regarding the five coupled equations~\eqref{eq:renorm-polaritons}, which microscopically describe polariton states down to the very strong coupling regime, is that it is straightforward to show formally that, in the regime of perturbative light–matter coupling, i.e., for tightly bound excitons, when the Rabi couplings $\Omega_\sigma$ are much smaller than the Rydberg energies, $\Omega_\sigma \ll R_{X\sigma}$, one recovers the polariton energies obtained by diagonalizing a coupled oscillator model:
\begin{equation}
\label{eq:5-coupled oscillator model}
    \begin{pmatrix}
        E_{hh,1s,+} &0 &0 &0 & \Omega_{hh}/\sqrt{2}\\
        0 & E_{hh,1s,-} & 0&0 & \Omega_{hh}/\sqrt{2}\\
        0 & 0& E_{lh,1s,+} &0 & \Omega_{lh}/\sqrt{2}\\
        0 & 0& 0& E_{lh,1s,-} & \Omega_{lh}/\sqrt{2}\\
        \Omega_{hh}/\sqrt{2} & \Omega_{hh}/\sqrt{2} &\Omega_{lh}/\sqrt{2} & \Omega_{lh}/\sqrt{2} & \nu
    \end{pmatrix} \, .
\end{equation}
This model simplifies further at zero magnetic field because of the absence of exciton Zeeman splitting, thus reducing to the diagonalisation of an effective 3 coupled oscillator model:
\begin{equation}
\label{eq:3-coupled oscillator model}
    \begin{pmatrix}
        E_{hh,1s}^{(B=0)} &0 & \Omega_{hh} & 0 & 0 \\
        0 & E_{lh,1s}^{(B=0)}& \Omega_{lh} & 0 & 0\\
        \Omega_{hh} & \Omega_{lh}& \nu & 0 & 0\\
        & 0 & 0 & E_{hh,1s}^{(B=0)} & 0\\
        & 0 & 0 & 0 & E_{lh,1s}^{(B=0)}
    \end{pmatrix} \, .
\end{equation}
We use this last model to fit the experimental energies of the first two polariton states in Fig.~\ref{fig:polariton-spectra-MF} at zero magnetic field to obtain the two values of the Rabi couplings $\Omega_\sigma$ listed in Tab.~\ref{tab:hh-lh_summary}. In particular, we get $\Omega_{hh} = 5.41$~meV and $\Omega_{lh} = 3.70$~meV. As noted in the main text, angular momentum algebra requires $\lambda_{lh}=\lambda_{hh}/\sqrt{3}\simeq0.58\lambda_{hh}$~\cite{ivchenko2005optical}. By using the relation~\eqref{eq:Rabi-c} between $\lambda_\sigma$ and $\Omega_\sigma$ and $a_{X,hh}/a_{X,lh}=1.28$, one obtains the following ratio for $\lambda_{lh}=0.54\lambda_{hh}$, consistent with the estimate from band theory.

Note that, in order to predict the behavior of the first three polariton states at arbitrary magnetic fields (and of the fourth state at moderate magnetic fields), it is sufficient to extract the values of $\Omega_\sigma$ from the experimental data uniquely at zero magnetic field.
However, we can apply a similar strategy to extract the evolution of the Rabi splittings from the experimental data and compare it with the theoretically expected behavior based on the exciton oscillator strength. We extend the analysis to also include the $hh, 2s$ state, in addition to the $hh$ and $lh$ $1s$ states.
We  extract the experimental values of $\Omega_{\sigma,ns}$ 
 by fitting an extended coupled oscillator model to the dispersion of the first three polariton states, labeled as lower polariton ($LP$), first ($UP_1$) and second upper polariton ($UP_2$) in Fig.~\ref{fig:polariton-spectra-MF} of the main text. These polariton states interpolate, as a function of the detuning, i.e., the waveguide momentum $\kappa_x$, between the following energies
\begin{center}
\begin{tabular}{llcl}
    $LP$: &  $\nu$ & $\leftrightarrow$ & $E_{hh, 1s,\pm}$\\
    $UP_1$: & $E_{hh, 1s,\pm}$ & $\leftrightarrow$ & $E_{lh, 1s,+}$\\
    $UP_2$: & $E_{lh, 1s,-}$ & $\leftrightarrow$ & $E_{hh, 2s,+}$\, .
\end{tabular}
\end{center}
where the energies $E_{hh, 1s,+} \simeq E_{hh, 1s,-}$ because of the negligible Zeeman splitting and where we are neglecting the $UP$ state that interpolates between $E_{lh, 1s,+}$ and $E_{lh, 1s,-}$ because it has negligible oscillator strength and can be measured in experiments only at very high magnetic fields. 
The energy dispersion of the three polariton states $LP$, $UP_1$, and $UP_2$ are then fitted (for different values of $\kappa_x$ around resonance) with the following generalized 7 coupled oscillator model, 
\begin{equation}
\label{eq:7-coupled oscillator model}
    \begin{pmatrix}
        E_{hh,1s,+} & 0 & 0 & 0 & 0 & 0 & \Omega_{hh,1s}/\sqrt{2}\\
        0 & E_{hh,1s,-} & 0 & 0 & 0 & 0 &   \Omega_{hh,1s}/\sqrt{2}\\
        0 & 0 & E_{lh,1s,+} & 0 & 0 & 0 & \Omega_{lh,1s}/\sqrt{2}\\
        0 & 0 & 0 & E_{lh,1s,-} & 0 & 0 & \Omega_{lh,1s}/\sqrt{2}\\
        0 & 0 & 0 & 0 & E_{hh,2s,+} & 0 & \Omega_{hh,2s}/\sqrt{2}\\
        0 & 0 & 0 & 0 & 0 & E_{hh,2s,-} & \Omega_{hh,2s}/\sqrt{2}\\
        \Omega_{hh},1s/\sqrt{2} & \Omega_{hh,1s}/\sqrt{2} &\Omega_{lh,1s}/\sqrt{2} & \Omega_{lh,1s}/\sqrt{2} & \Omega_{hh,2s}/\sqrt{2} & \Omega_{hh,2s}/\sqrt{2} & \nu
    \end{pmatrix} \, .
\end{equation}
where the exciton energies, $E_{\sigma,ns,\pm}$ are fixed to the values extracted from reflectance measurements in weak coupling, while $\Omega_{\sigma,ns}$ are fitting parameters. In this way, we extract the values of $\Omega_{\sigma,ns}$ for three values of the magnetic field, $B=2, 4, 7$~T.  We fit the polariton dispersions around the resonances between the exciton states and the waveguide state TE$_{1,+}$ and TE$_{1,-}$ separately, because they lead to slightly different values of the Rabi splittings.  
For the highest value of the magnetic field $B=9$~T, we decide to fit only the values of $\Omega_{hh,1s}$ and $\Omega_{lh,1s}$ with the 5 coupled oscillator model~\eqref{eq:5-coupled oscillator model}. This is because the exciton $hh, 2s$ states move far away up in energy and also because additional avoided crossing in between the $lh, 1s, +$ state and the $hh, 2s$ appear possibly due to exciton 
dark states which have gained oscillator strength due to magnetic field induced mixing with bright excitons---see accompanying paper~\cite{long-paper} for more details.

\begin{figure}
    \centering
    \includegraphics[width=\linewidth]{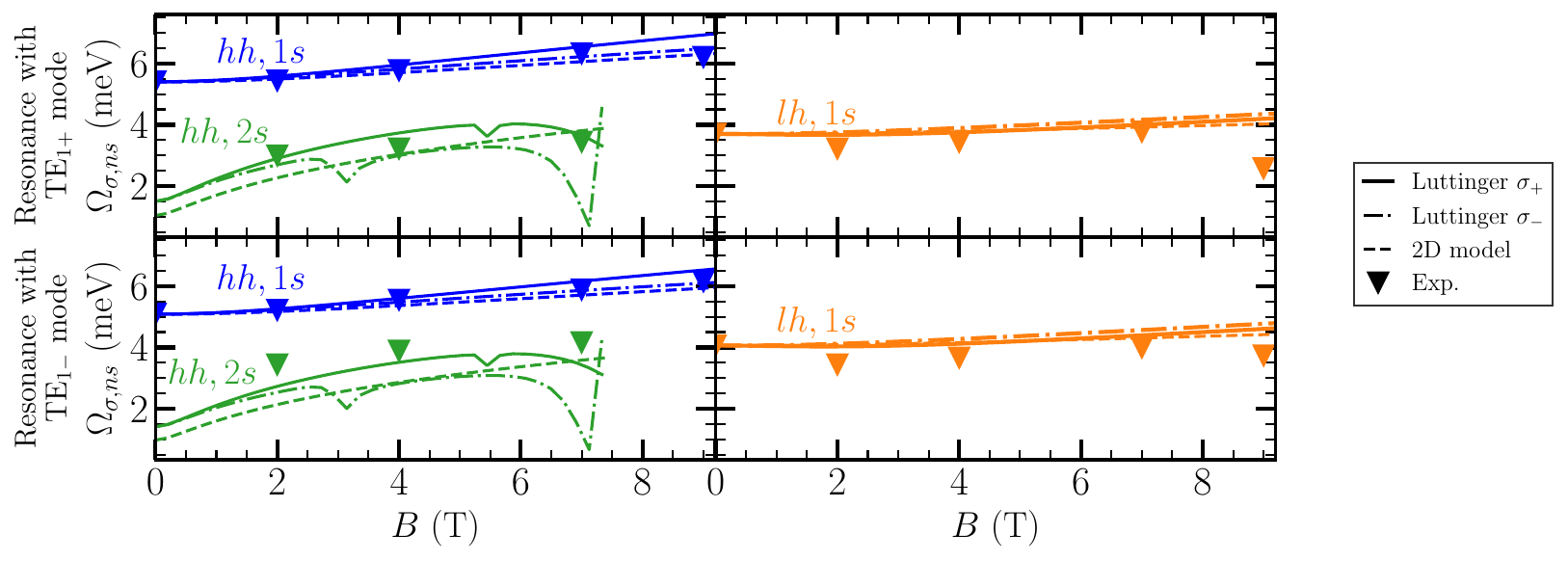}\caption{ 
    Comparison between the Rabi couplings $\Omega_{\sigma,ns}$ extracted from experiments (symbols) and the theoretical predictions (lines) derived from the exciton oscillator strength. Dashed lines correspond to the predictions within the approximate 2D model~\eqref{eq:2Dtheory-Rabi}. Solid (dotted) lines correspond to the predictions~\eqref{eq:3Dtheory-Rabi} from the Luttinger model, which yields two values for the Rabi splittings corresponding to the two light polarizations, $+$ and $-$, that are very close to each other.
    Top and bottom panels refer to the case of Rabi couplings extracted by fitting the polariton modes close to resonance with either the TE$_{1,+}$ (top) or TE$_{1,-}$ (bottom).
    }
\label{fig:osc-str-comparison}
\end{figure}

The Rabi couplings $\Omega_{\sigma, ns}$ extracted from the experiments are plotted in Fig.~\ref{fig:osc-str-comparison} together with the theoretical predictions based on the exciton oscillator strength. 
These are obtained from the expression of the exciton wave function for the $\sigma, ns$ state at zero electron-hole separation, $\varphi_{\sigma,ns}(0)$ and by appropriately rescaling it by the $B=0$ values of the $1s$ state, $\varphi_{\sigma,1s}^{(B=0)}(0)$ and $\Omega_\sigma$~\cite{Laird_PRB2022}:
\begin{equation}
\label{eq:2Dtheory-Rabi}
    \Omega_{\sigma,ns}=\Omega_{\sigma} \frac{|\varphi_{\sigma,ns}(\rho=0)|}{|\varphi^{(B=0)}_{\sigma,1s}(\rho=0)|}\, .
\end{equation}
Note that, in the effective 2D exciton model~\eqref{eq:2D-exciton}, the obtained values of $\Omega_{\sigma,ns}$ at finite magnetic field are, by definition, independent of the light polarization $\pm$, since the exciton wave function is independent of it. In contrast, in the Luttinger model~\eqref{eq:Luttinger-exciton}, one obtains in principle different values of $\Omega_{\sigma,ns}$ for $+$ and $-$ light polarization:
\begin{subequations}
\label{eq:3Dtheory-Rabi}
\begin{align}
    \Omega_{hh,ns,\pm}=&\Omega_{hh}\frac{\left|\int_{-\infty}^{\infty} dz \varphi_{J_h=\pm\frac{3}{2},\ell=0, ns}^{J_z=\pm1, J_e=\mp\frac{1}{2}}(\rho=0,z,z)\right|}{\left|\int_{-\infty}^{\infty} dz \varphi_{J_h=\pm\frac{3}{2},\ell=0, 1s}^{J_z=\pm1, J_e=\mp\frac{1}{2} (B=0)}(\rho=0,z,z)\right|}\\
    \Omega_{lh, ns, \pm}=&\Omega_{lh}\frac{\left|\int_{-\infty}^{\infty}dz \varphi_{J_h=\pm\frac{1}{2},\ell=0,ns}^{J_z=\pm 1, J_e=\pm\frac{1}{2}}(\rho=0,z,z)\right|}{\left|\int_{-\infty}^{\infty}dz \varphi_{J_h=\pm\frac{1}{2},\ell=0,1s}^{J_z=\pm1, J_e=\pm\frac{1}{2} (B=0)}(\rho=0,z,z)\right|},
\end{align}    
\end{subequations}
where $\varphi_{J_h,\ell=0,ns}^{J_z,J_e}(\rho=0,z,z)$ is the $ns$-wave $J_h=\frac{3}{2}$ ($hh$) or $J_h=\frac{1}{2}$ ($lh$) component of the multicomponent exciton at zero relative electron-hole distance and at finite magnetic field.
Note, however, that the Rabi splittings obtained within the Luttinger model for the $+$ and $-$ polarizations are nearly identical, thereby justifying the approximation employed in the 2D modeling of polaritons.
Indeed, the magnetic-field dependence of the oscillator strength predicted by the 2D model closely follows that obtained from the Luttinger model.
The comparison between experimental results and theoretical predictions in Fig.~\ref{fig:osc-str-comparison} shows very good agreement.
Note that, for the $lh, 1s$ state, the Luttinger model predicts a slight increase of the oscillator strength with magnetic field, in agreement with the 2D model, as a consequence of the enhancement of the binding energy. Notably, in the Luttinger model this occurs despite the inclusion of mixing effects with dark exciton states (see Ref.~\cite{long-paper}).

We compare in Fig.~\ref{fig:UP-diamagnetic-shift} of the main text the experimental results for the energy shift  $\Delta E = E^{(B)} - E^{(B=0)}$ of the $UP_1$ polariton state with the non-perturbative results obtained solving Eq.~\eqref{eq:renorm-polaritons}, as well as the perturbative ones coming from the diagonalization of the coupled oscillator model~\eqref{eq:5-coupled oscillator model}. For the latter we use the Rabi splittings  $\Omega_{\sigma,ns}$ calculated using Eq.~\eqref{eq:2Dtheory-Rabi} and plotted in Fig.~\ref{fig:osc-str-comparison}.

\begin{figure}
    \centering
    \includegraphics[width=\linewidth]{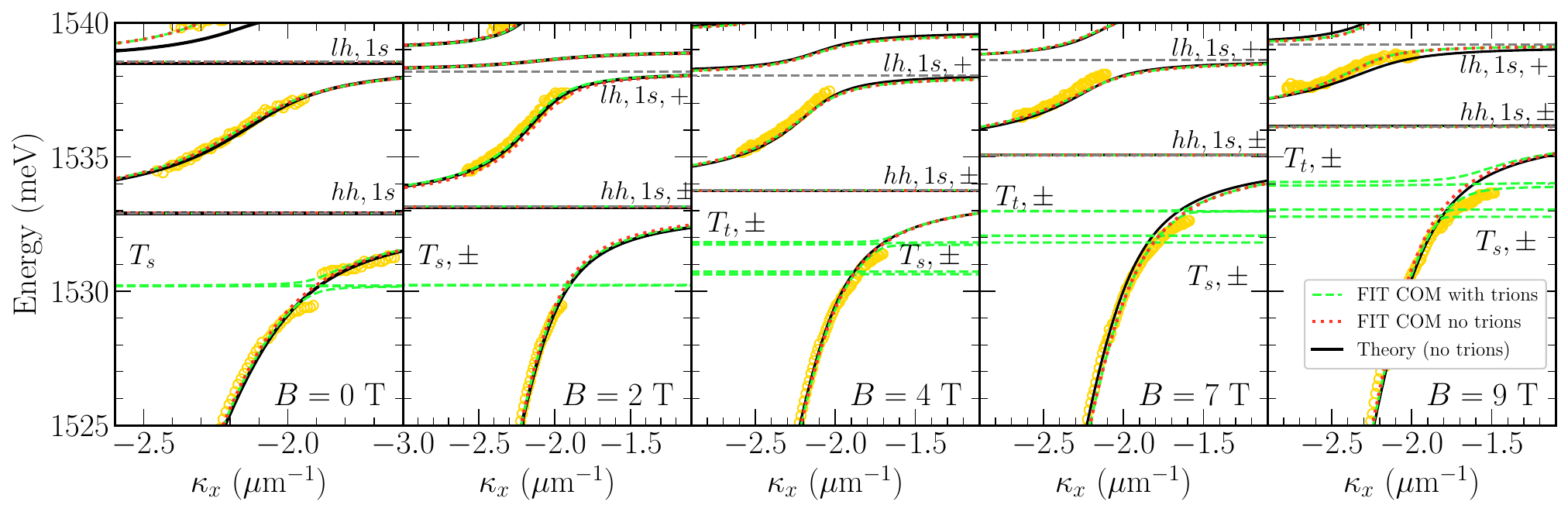}
    \caption{{Energy dispersion of the lower ($LP$) and first upper ($UP_1$) polariton branches at different magnetic fields as a function of the in-plane momentum $\kappa_x$. Experimental data are shown as yellow circles (same as in Fig.~\ref{fig:polariton-spectra-MF} of the main text and Fig.~\ref{fig:additional-states-weak-coupling}). Solid black lines represent the theoretical results obtained from the microscopic 2D polariton model described in this Supplementary Material and in the main text, which excludes contributions from trion states. Dotted red and dashed green lines correspond to the results of a coupled oscillator model fitting procedure that either excludes or includes trion states, respectively. The extracted Rabi couplings obtained from fitting the $LP$ state while including the trion resonances are: $\Omega_{T_s}^{(B=0)}=0.62$~meV, $\Omega_{T_s}^{(B=2~\text{T})}=0.04$~meV, $\Omega_{T_s}^{(B=4~\text{T})}=0.05$~meV, $\Omega_{T_s}^{(B=7~\text{T})}=0$~meV,   $\Omega_{T_s}^{(B=9~\text{T})}=0.2$~meV}, $\Omega_{T_t}^{(B=4~\text{T})}=0.47$~meV,  $\Omega_{T_t}^{(B=7~\text{T})}=0.32$~meV, and $\Omega_{T_t}^{(B=9~\text{T})}=1.18$~meV  (see text). } 
\label{fig:LP-anticrossing_checkUP}
\end{figure}
%
\section{Trion states and lower polariton dispersion}
\label{app:Lower-polariton}
In this section, we justify the neglect of trion states below the $hh, 1s$ exciton in our polariton model. We show that including these states does not affect the properties of the upper polariton branches, which are the primary focus of this work, as they carry signatures of the very strong coupling regime. 

We extend the coupled oscillator models~\eqref{eq:5-coupled oscillator model} and~\eqref{eq:7-coupled oscillator model} by incorporating the trion states observed in the weak-coupling regime. This is achieved by adding the following terms to Eqs.~\eqref{eq:5-coupled oscillator model} and~\eqref{eq:7-coupled oscillator model}:
\begin{equation}
\label{eq:reduced-coupled oscillator model}
\begin{pmatrix}
    E_{T_{s,+}} & 0 & 0 & 0 & \cdots & \Omega_{T_s}/\sqrt{2} \\
    0 & E_{T_{s,-}} & 0 & 0 & \cdots & \Omega_{T_s}/\sqrt{2} \\
    0 & 0 & E_{T_{t,+}} & 0 & \cdots & \Omega_{T_t}/\sqrt{2} \\
    0 & 0 & 0 & E_{T_{t,-}} & \cdots & \Omega_{T_t}/\sqrt{2} \\
    \vdots & \vdots & \vdots & \vdots & \ddots & \vdots \\
    \Omega_{T_s}/\sqrt{2} & \Omega_{T_s}/\sqrt{2} & \Omega_{T_t}/\sqrt{2} & \Omega_{T_t}/\sqrt{2} & \cdots & \nu
\end{pmatrix}\, .
\end{equation}
The choice between the two cases, ~\eqref{eq:5-coupled oscillator model} and~\eqref{eq:7-coupled oscillator model}, depends on whether the $hh,2s$ exciton is included in the fitting procedure ($B = 2, 4, 7$~T) or excluded ($B = 0, 9$~T).
The triplet trion states are included in the coupled oscillator model only at $B = 4, 7, 9$~T. The trion ($E_{T_t,\pm}$ and $E_{T_s,\pm}$), exciton ($E_{hh,1s,\pm}$, $E_{hh,2s,\pm}$, and $E_{lh,1s,\pm}$), and waveguide ($\nu$) energies are fixed to the experimental values. In addition, the exciton Rabi splittings $\Omega_{\sigma,ns}$ are fixed to the previously reported values (see Table~\ref{tab:hh-lh_summary} and Fig.~\ref{fig:osc-str-comparison}), while the trion Rabi splittings $\Omega_{T_t}$ and $\Omega_{T_s}$ are extracted by fitting the polariton dispersions. The results are shown in Fig.~\ref{fig:LP-anticrossing_checkUP}.
Here, we compare the polariton branch dispersions obtained in this way with those obtained without including trion states---i.e., from Eq.~\eqref{eq:5-coupled oscillator model} at $B = 0, 9$~T and Eq.~\eqref{eq:7-coupled oscillator model} at $B = 2, 4, 7$~T. We also compare these results with those from our microscopic 2D polariton model, which similarly neglects trion contributions.

We find that including the trion states in the coupled oscillator model does not affect the energy dispersion of the $UP_1$ mode at any magnetic field, compared with the dispersion obtained previously without trions. Although the differences between the coupled oscillator model and the microscopic 2D model appear small on the scale shown in Fig.~\ref{fig:LP-anticrossing_checkUP}, the $UP_1$ energy shift can be accurately captured only by the latter, as discussed in the main text (see Fig.~\ref{fig:UP-diamagnetic-shift}). This is further evident from the analysis of the $UP_1$ composition at zero magnetic field, presented in the following section, which reveals the contribution of the $hh$ continuum. Such contributions cannot be included in a coupled oscillator model, which accounts for the coupling to light only perturbatively. 

Regarding the experimental $LP$ dispersion extracted from the reflectance spectra, we observe a weak anticrossing around the energy of the singlet trion state at $B = 0$~T and at $B = 2$~T, although in the latter case the upper $LP$ branch cannot be resolved. At $B = 0$~T, fitting the coupled oscillator model gives a Rabi coupling $\Omega_{T_s}^{(B=0~\text{T})} = 0.62$~meV.
As shown in Fig.~\ref{fig:additional-states-strong-coupling}, at finite magnetic fields the avoided crossing with the singlet trion $T_s$ is suppressed, while the coupling of the triplet trion $T_t$ state to light is enhanced. The presence of trion states below the $hh, 1s$ exciton explains the deviation of the experimental $LP$ data from the predictions of the microscopic 2D model, which includes only neutral excitons.

\section{Composition analysis of the upper polariton states}
\label{sec:upper-polariton-analysis}
In this section, we study the properties of the upper polariton states, including their exciton and photon fractions and the contributions from different exciton Rydberg states, as a function of magnetic field and photon–exciton detuning.
We show that the presence of a $lh, 1s$ exciton state between the $hh, 1s$ and $hh, 2s$ states, arising from the large QW width, strongly influences the composition and properties of the $UP_1$ and $UP_2$ states. This leads to marked differences compared to narrow-width QWs, where the $lh, 1s$ exciton lies well above the $hh$ Rydberg series, as analyzed in Ref.~\cite{Laird_PRB2022}.
The analysis presented in this section provides a detailed insight into the origin of the deviations between the nonperturbative microscopic theory and the perturbative coupled oscillator model in predicting the magnetic-field-induced energy shifts of the $UP_1$ and $UP_2$ states, shown in Fig.~\ref{fig:UP-diamagnetic-shift} of the main text. These deviations constitute signatures of the very strong coupling regime.

The photon and exciton fractions of a given polariton state $\ket{P}$~\eqref{eq:pol-state} are given, respectively, by:
\begin{align}
\label{eq:fractions}
    &|\gamma|^2 & |X_{\sigma,\pm}|^2 &= \int d\rhov|\varphi_{\sigma,\pm}(\rhov)|^2\, ,
\end{align}
and normalization~\eqref{eq:normalization-P} requires that
\begin{equation}
    \sum_{\sigma,\pm}|X_{\sigma,\pm}|^2
      +|\gamma|^2=1\, .
\end{equation}
Furthermore, another useful quantity for describing polariton states is the overlap of a given polariton state with the corresponding bare exciton state in the weak-coupling regime. This quantity, which we denote as the bare-exciton overlap, is defined as:
\begin{equation}
\label{eq:bare-exciton-overlap}
    P_{\sigma,ns,\pm} = \frac{|\bra{P} \ket{X_{\sigma,ns}}|^2}{\int d\rhov |\varphi_{\sigma, \pm}(\rho)|^2} = \frac{|\int d\rhov \varphi_{\sigma,ns}^{*}(\rho) \varphi_{\sigma, \pm}(\rho)|^2}{\int d\rhov |\varphi_{\sigma, \pm}(\rho)|^2}\, .
\end{equation}
In this expression, $\ket{P=LP, UP_1, \dots}$ denotes the polariton state defined in Eq.~\eqref{eq:pol-state}, whose electron–hole relative wave function is $\varphi_{\sigma,\pm}(\rho)$ and which satisfies the coupled Eqs.~\eqref{eq:renorm-polaritons} together with the photon amplitude $\gamma$. By contrast, $\ket{X_{\sigma,ns}}$ denotes the bare $ns$ exciton state, whose wave function $\varphi_{\sigma,ns}$ satisfies the \sch equation~\eqref{eq:2D-exciton}---we recall that, within the 2D model, the exciton energy is polarization dependent, $E_{\sigma,ns,\pm}$, whereas the corresponding wave function is polarization independent.
Because the bare exciton states form a complete basis set, the bare-exciton overlap~\eqref{eq:bare-exciton-overlap} is normalized to unity:
\begin{equation*}
    \sum_{n}^{\text{bound}+\text{unbound}}  P_{\sigma,ns,\pm}= 1\, ,
\end{equation*}
where the sum over $n$ runs over both bound and unbound bare exciton states. 
Note that, at finite magnetic field, bound exciton states are defined such that their energy, measured with respect to the corresponding Landau levels of a free electron–hole pair, is negative~\cite{Laird_PRB2022}, i.e.:
\begin{equation}
    \Delta E_{\sigma,ns, \pm} = E_{\sigma, ns,\pm}  - \left[
    E_{g\sigma} + (2n-1) \omega_{c\sigma} \pm g_{\sigma} \mu_B B\right] < 0 \, . 
\end{equation}

\begin{figure}
\centering
\includegraphics[width=\linewidth]{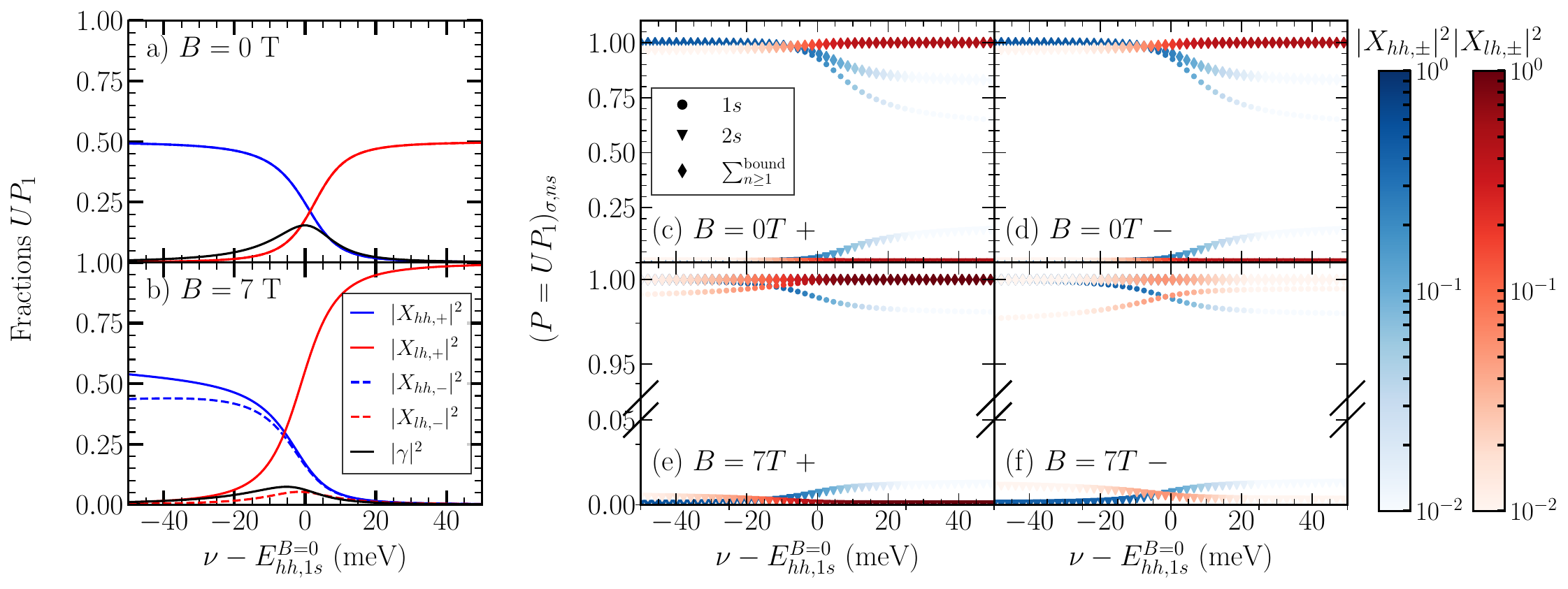}
    \caption{{Left panels: exciton and photon fractions~\eqref{eq:fractions} of the $UP_{1}$ state as a function of the photon-exciton detuning at (a) $B=0$~T and (b) $B=7$~T. Right panels: bare-exciton overlap~\eqref{eq:bare-exciton-overlap} of the $UP_{1}$ state as a function of the photon-exciton detuning at (a) $B=0$~T at $B=0$~T (c,d) and $B=7$~T (e,f), with $+$ polarization components in panels (c,e) and $-$ in panels (d,f). Dots correspond to the $1s$ contribution ($P_{\sigma,1s,\pm}$), while triangles denote the $2s$ contribution ($P_{\sigma,2s,\pm}$). Diamonds represent the total contribution from all bound states, $\sum_{n=1}^{\text{bound}} P_{\sigma,ns,\pm} = 1 - \sum_{n=1}^{\text{unbound}} P_{\sigma,ns,\pm}$; thus, deviations from unity indicate contributions from the continuum. Same color scheme as the left panels: $\sigma=hh$ in blue and $\sigma=lh$ in red. In the right panel, the color intensity of each data point is weighted by the exciton fraction.}}
\label{fig:UP1-fractions-composition}
\end{figure}

In Fig.~\ref{fig:UP1-fractions-composition}, we plot the photon and exciton fractions, as well as the bare-exciton overlap of the $UP_1$ state as a function of the photon–exciton detuning and for different values of the magnetic field.
We recall that the energy of the $UP_1$ state interpolates, as a function of detuning, between  $E_{hh,1s,\pm}$ (remember that $E_{hh,1s,+} \simeq E_{hh,1s,-} $) at negative detuning  and $E_{lh,1s,+}$ at positive detuning. Accordingly, the $hh$ and $lh$ fractions shown in the left panels follow this trend. The photon fraction, by contrast, is largest in the crossover region near resonance and decreases overall with increasing magnetic field, because the $E_{hh,1s,\pm}$ level comes closer to $E_{lh,1s,+}$.
{The bare-exciton overlap of the $UP_1$ state, shown in the right panels of Fig.~\ref{fig:UP1-fractions-composition}, indicates that the largest contributions to its matter-wavefunction component come from the $hh, 1s$ and $lh, 1s$ exciton states. The difference between $B=0$~T and $B=7$~T is that, at $B=0$~T, there is a finite contribution from the $hh,2s$ state, the excited Rydberg states, as well as the continuum of unbound $hh$ states. This continuum contribution increases toward positive detuning, even though the overall $hh$ fraction decreases in this regime. By contrast, it is significantly reduced at finite magnetic field.} 
This behavior arises because the $UP_1$ energy shifts toward that of the $lh,1s$ state at positive detuning. At zero magnetic field, this state lies closer to the $hh,2s$ level and therefore has a larger overlap with it than at finite magnetic field.

This analysis implies that, at finite magnetic field, light predominantly hybridizes the $hh,1s$ and $lh,1s$ exciton states in the $UP_1$ mode, whereas at zero magnetic field the hybridization also involves the $hh,2s$ state, higher Rydberg states, and the continuum of unbound states.
Furthermore, this explains why the energy shift $\Delta E = E^{(B)} - E^{(B=0)}$ of the $UP_1$ polariton state predicted by the microscopic 2D model does not deviate significantly from that obtained using the coupled oscillator model (see Fig.~\ref{fig:UP-diamagnetic-shift}), in contrast to the case of narrow QWs~\cite{Laird_PRB2022}. The deviations are nevertheless large enough to allow experimental detection of light-induced modifications of the exciton wave function through the polariton energy shift.

\begin{figure}
    \centering
    \includegraphics[width=\linewidth]{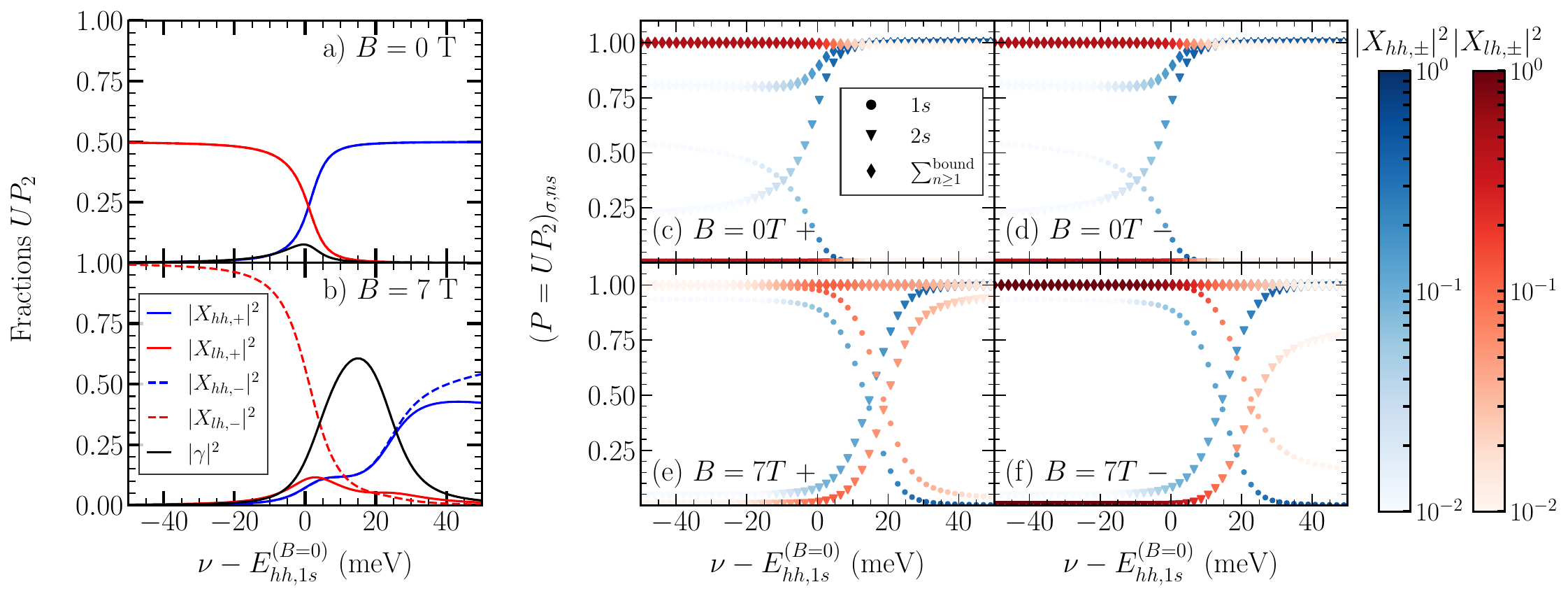}
    \caption{Left panels: exciton and photon fractions~\eqref{eq:fractions} of the $UP_{2}$ state as a function of the photon-exciton detuning at (a) $B=0$~T and (b) $B=7$~T. Right panels: bare-exciton overlap~\eqref{eq:bare-exciton-overlap} of the $UP_{2}$ state as a function of the photon-exciton detuning at (a) $B=0$~T at $B=0$~T (c,d) and $B=7$~T (e,f), with $+$ polarization components in panels (c,e) and $-$ in panels (d,f). Dots correspond to the $1s$ contribution ($P_{\sigma,1s,\pm}$), while triangles denote the $2s$ contribution ($P_{\sigma,2s,\pm}$). Diamonds represent the total contribution from all bound states, $\sum_{n=1}^{\text{bound}} P_{\sigma,ns,\pm} = 1 - \sum_{n=1}^{\text{unbound}} P_{\sigma,ns,\pm}$; thus, deviations from unity indicate contributions from the continuum. Same color scheme as the left panels: $\sigma=hh$ in blue and $\sigma=lh$ in red. In the right panel, the color intensity of each data point is weighted by the exciton fraction.}
\label{fig:UP2-fractions-composition}
\end{figure}

\begin{figure}
    \centering
\includegraphics[width=.7\linewidth]{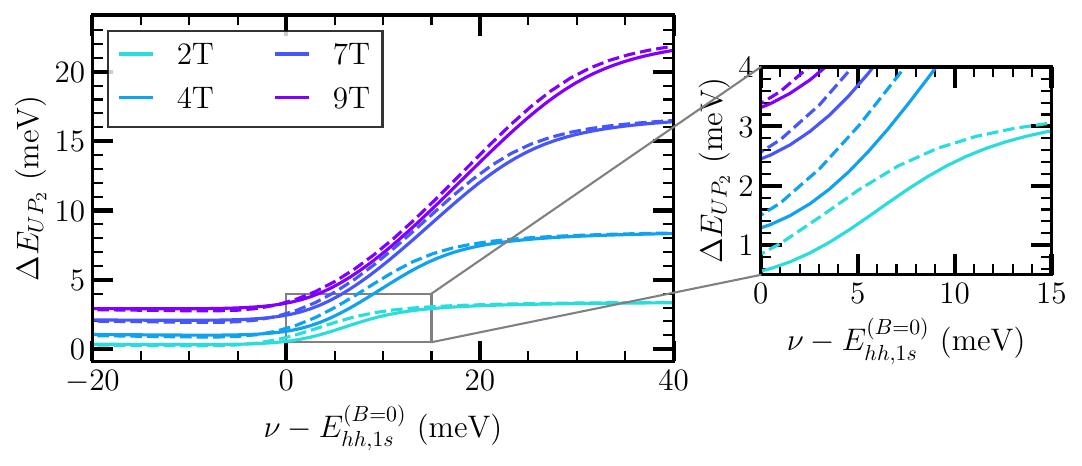}
    \caption{{Energy shift, $\Delta E = E^{(B)} - E^{(B=0)}$, of the polariton state $UP_2$ at different magnetic fields. Solid lines show the results from the 2D polariton model, while dashed lines represent the perturbative results obtained from the coupled oscillator model~\eqref{eq:7-coupled oscillator model}.
    }}
\label{fig:UP2-energy-shift-MF}
\end{figure}

We now examine the properties of the $UP_2$ polariton state. In particular, we show that the presence of the $lh,1s$ state between the $hh$ Rydberg states also reshapes the hybridization of heavy-hole excitons within $UP_2$.
The energy of the $UP_2$ state interpolates between $E_{lh,1s,-}$ and  $E_{hh,2s,-}$ as the photon energy grows. In the left panel of Fig.~\ref{fig:UP2-fractions-composition}, we examine the photon and exciton fractions at zero and finite magnetic field. 
At zero magnetic field, the small energy separation between the $lh,1s$ and $hh,2s$ states keeps the photon fraction $|\gamma|^2$ of $UP_2$ low, while at finite magnetic field, the larger separation increases $|\gamma|^2$. 
As expected, the dominant contributions to the bare-exciton overlap of the $UP_2$ state come from the $hh,1s$ exciton (dominant at negative detuning) and the $hh,2s$ exciton (dominant at positive detuning), both at zero and finite magnetic fields.
Similarly to  what happens for the $UP_1$, the main difference between zero and finite magnetic field is that, at $B=0$, there is a significant contribution of the continuum $hh$ states, whereas this contribution disappears again finite $B$. 
However, unlike $UP_1$, the $UP_2$ state exhibits a significant contribution from the $lh, 2s$ state at finite $B$ for positive detuning.
To conclude, Fig.~\ref{fig:UP2-energy-shift-MF} shows the energy shift of the $UP_2$ state at different magnetic fields, comparing the nonperturbative results with those obtained from the coupled oscillator model. 
These deviations between the two results remain overall small, because the contribution of the $hh$ continuum is suppressed at finite magnetic field, while the bare-exciton overlap of $UP_2$ with the $lh, 2s$ state is largest when $|X_{lh,\pm}|^2$ is minimal.


\end{document}